\documentclass[twocolumn,amsmath,amssymb,prd]{revtex4}

\def\al{\alpha}
\def\be{\beta}

\def\de{\delta}
\def\ep{\epsilon}

\def\th{\theta}

\def\ta{\tau}

\def\ph{\phi}

\def\ch{\chi}

\def\om{\omega}

\def\De{\Delta}

\def\ab{{\al\be}}
\def\cA{{\cal A}}

\def\fr#1#2{{{#1} \over {#2}}}

\def\half{{\textstyle{1\over 2}}}
\def\lsim{\mathrel{\rlap{\lower4pt\hbox{\hskip1pt$\sim$}}
    \raise1pt\hbox{$<$}}}
\def\gsim{\mathrel{\rlap{\lower4pt\hbox{\hskip1pt$\sim$}}
    \raise1pt\hbox{$>$}}}
\def\Re{\hbox{Re}\,}
\def\Im{\hbox{Im}\,}

\def\etal {{\it et al.}}
\newcommand{\beq}{\begin{equation}}
\newcommand{\eeq}{\end{equation}}
\newcommand{\bea}{\begin{eqnarray}}
\newcommand{\eea}{\end{eqnarray}}
\newcommand{\bse}{\begin{subequations}}
\newcommand{\ese}{\end{subequations}}
\newcommand{\rf}[1]{(\ref{#1})}

\def\to{\rightarrow}

\def\mix{\leftrightarrow}
\def\nub{\bar\nu}

\def\heff{h_{\rm eff}}
\def\gt{\widetilde g} 
\def\Ht{\widetilde H} 
\def\aL{(a_L)}
\def\cL{(c_L)}
\def\nub{\bar\nu}
\def\heff{h_{\rm eff}}
\def\As#1{({\cal A}_s)_{#1}}
\def\Ac#1{({\cal A}_c)_{#1}}
\def\Bs#1{({\cal B}_s)_{#1}}
\def\Bc#1{({\cal B}_c)_{#1}}
\def\C#1{({\cal C})_{#1}}
\def\Asn#1#2{({\cal A}^{(#1)}_s)_{#2}}
\def\Acn#1#2{({\cal A}^{(#1)}_c)_{#2}}
\def\Bsn#1#2{({\cal B}^{(#1)}_s)_{#2}}
\def\Bcn#1#2{({\cal B}^{(#1)}_c)_{#2}}
\def\Dsn#1#2{({\cal D}^{(#1)}_s)_{#2}}
\def\Dcn#1#2{({\cal D}^{(#1)}_c)_{#2}}
\def\Fsn#1#2{({\cal F}^{(#1)}_s)_{#2}}
\def\Fcn#1#2{({\cal F}^{(#1)}_c)_{#2}}
\def\Cn#1#2{({\cal C}^{(#1)})_{#2}}
\def\PAsn#1#2{(P_{{\cal A}_s}^{(#1)})_{#2}}
\def\PAcn#1#2{(P_{{\cal A}_c}^{(#1)})_{#2}}
\def\PBsn#1#2{(P_{{\cal B}_s}^{(#1)})_{#2}}
\def\PBcn#1#2{(P_{{\cal B}_c}^{(#1)})_{#2}}
\def\PDsn#1#2{(P_{{\cal D}_s}^{(#1)})_{#2}}
\def\PDcn#1#2{(P_{{\cal D}_c}^{(#1)})_{#2}}
\def\PFsn#1#2{(P_{{\cal F}_s}^{(#1)})_{#2}}
\def\PFcn#1#2{(P_{{\cal F}_c}^{(#1)})_{#2}}
\def\PCn#1#2{(P_{\cal C}^{(#1)})_{#2}}
\def\M#1#2#3{({\cal M}^{(#1)}_{#2})_{#3}}
\def\H#1#2{{\cal H}^{(#1)}_{#2}}
\def\T#1#2{\ta^{(#1)}_{#2}}
\def\S#1#2{S^{(#1)}_{#2}}
\def\aR{(a_R)}
\def\cR{(c_R)}
\def\AL{(\widetilde a_L)}
\def\AR{(\widetilde a_R)}
\def\CL{(\widetilde c_L)}
\def\CR{(\widetilde c_R)}
\def\gT{\widetilde{\widetilde g}{}} 
\def\HT{\widetilde{\widetilde H}{}} 
\def\Nh{{\hat N}}
\def\Ep{{\hat {\cal E}_+}}

\def\PASN#1{P_{{\cal A}_s}^{(#1)}}
\def\PACN#1{P_{{\cal A}_c}^{(#1)}}
\def\PBSN#1{P_{{\cal B}_s}^{(#1)}}
\def\PBCN#1{P_{{\cal B}_c}^{(#1)}}
 
\def\ol#1{\overline{#1}}

\begin{document}

\title{Perturbative Lorentz and CPT violation 
for neutrino and antineutrino oscillations}

\author{Jorge S.\ D\'iaz,$^1$
V.\ Alan Kosteleck\'y,$^1$ 
and Matthew Mewes$^2$}

\affiliation{
$^1$Physics Department, Indiana University, 
Bloomington, IN 47405, U.S.A.\\
$^2$Department of Physics and Astronomy, Swarthmore College, 
Swarthmore, PA 19081, U.S.A.}

\date{IUHET 528, August 2009; 
published as Phys.\ Rev.\ D {\bf 80}, 076007 (2009)}

\begin{abstract}
The effects of perturbative Lorentz and CPT violation 
on neutrino oscillations are studied.
Features include 
neutrino-antineutrino oscillations,
direction dependence,
and unconventional energy behavior.
Leading-order corrections 
arising from renormalizable operators are derived 
in the general three-flavor effective field theory.
The results are applied to neutrino-beam experiments 
with long baselines,
which offer excellent sensitivity to the accompanying effects.
Key signatures of Lorentz and CPT violation using neutrino beams
include sidereal variations in the oscillation probabilities
arising from the breakdown of rotational symmetry,
and CPT asymmetries comparing neutrino and antineutrino modes.
Attainable sensitivities 
to coefficients for Lorentz violation 
are estimated for several existing and future experiments.
\end{abstract}

\maketitle

\section{Introduction}

Experimental investigations of neutrino properties 
have provided crucial insights into particle physics
since the existence of neutrinos
was first proposed in 1930 by Pauli 
\cite{pauli}
to explain the spectrum of beta decay.
In recent years,
the confirmed observation of neutrino oscillations
has established the existence of physics 
beyond the minimal Standard Model (SM)
\cite{pdg}.
The interferometric nature of the oscillations
makes them highly sensitive to new physics,
including potential low-energy signals
that may originate in a fundamental theory 
unifying quantum physics and gravity 
at the Planck scale $m_P\simeq 10^{19}$ GeV.

In this work,
we investigate the experimental implications 
for neutrino oscillations 
of Lorentz and CPT violation, 
which is a promising category of Planck-scale signals
\cite{ks}.
The SM is known to provide a successful description 
of observed phenomena at energies well below $m_P$.
As a consequence,
the manifestation of Planck-scale effects
involving Lorentz and CPT violation
is expected to be well described 
at accessible energy scales
by an effective field theory containing the SM 
\cite{kp,reviews}.

The comprehensive effective field theory
describing general Lorentz violation
at attainable energies
is the Standard-Model Extension (SME)
\cite{ck,akgrav}.
It incorporates both the SM and General Relativity,
serving as a realistic theory for analyses of experimental data.
In the SME Lagrange density,
each Lorentz-violating term 
is an observer scalar density
constructed as the product of a Lorentz-violating operator
with a controlling coefficient.
Under mild assumptions,
CPT violation in effective field theory
is accompanied by Lorentz violation
\cite{owg},
so the SME also describes general breaking of CPT symmetry.
These ideas have triggered a wide variety of tests 
over the past decade
\cite{tables}.
Several experimental searches 
have been performed using neutrino oscillations,
yielding high sensitivities to SME coefficients 
for Lorentz and CPT violation
\cite{lsndlv,sklv,minoslv}.

Since both Lorentz-violating operators and mass terms 
can induce neutrino mixing,
one way to classify neutrino models with Lorentz and CPT violation
is in terms of their neutrino-mass content.
Three categories exist:
massless Lorentz-violating models, 
in which no neutrinos have mass;
hybrid Lorentz-violating models, 
with mass terms for a subset of neutrinos;
and
massive Lorentz-violating models, 
where all neutrinos have conventional masses.

In massless Lorentz-violating models,
all observed neutrino oscillations are attributed
to nonzero coefficients for Lorentz violation
rather than to masses.
Certain coefficients can combine via a Lorentz-seesaw mechanism
to produce pseudomasses that mimic
the behavior of mass terms
for a range of neutrino energies
\cite{km1}.
The prototypical example is the bicycle model
\cite{km2}
which uses two nonzero coefficients for Lorentz violation
to reproduce the expected behavior of atmospheric neutrinos.
This model agrees well with atmospheric data 
from the Super-Kamiokande experiment
\cite{sklv}.
However,
a combined analysis of neutrino data excludes 
both the bicycle model in its simplest form
and a five-parameter generalization 
\cite{bmw}.
Massless models may also predict sidereal signals
arising from the violation of rotation invariance
\cite{km3}.
The corresponding coefficients for Lorentz violation
have been constrained in experimental analyses
by the Liquid Scintillator Neutrino Detector
(LSND)
\cite{lsndlv}
and 
the Main Injector Neutrino Oscillation Search (MINOS)
\cite{minoslv}.
At present,
it is an interesting open challenge to construct 
a massless Lorentz-violating model
that is globally compatible with existing neutrino data. 

For hybrid Lorentz-violating models,
the tandem model 
\cite{kkt}
is the sole exemplar.
It is a three-parameter model
containing one neutrino mass,
one coefficient for CPT-even Lorentz violation, 
and one coefficient for CPT-odd Lorentz violation.
The model appears globally compatible 
with existing experimental data, 
including the LSND anomaly
\cite{lsndanomaly}.
The tandem model predicted a low-energy excess 
in the Mini Booster Neutrino Experiment (MiniBooNE) 
prior to its discovery,
although the observed excess is quantitatively greater
\cite{mbexcess}.
This success suggests that further theoretical investigations
of hybrid Lorentz-violating models would be of definite interest. 

Massive Lorentz-violating models 
are the primary focus of the present work.
Most existing data from neutrino oscillations
are consistent with oscillation phases 
proportional to the baseline $L$ 
and inversely proportional to the energy $E$.
This is conventionally interpreted as a consequence 
of mixing induced by a nondegenerate mass matrix.
In massive Lorentz-violating models,
the mixing due to mass is assumed to dominate
over that due to Lorentz violation.
Our goal here is to present a general study
of perturbative Lorentz and CPT violation
on mass-induced mixing,
valid over a wide range of $L$ and $E$.

The analysis presented here incorporates 
all coefficients for Lorentz violation
associated with quadratic operators of renormalizable dimension
in the neutrino sector
\cite{km1}.
Using notation reviewed in Sec.\ \ref{sec theory},
these coefficients are 
$\aL^\al_{ab}$, 
$\cL^\ab_{ab}$, 
$\gt^\ab_{a \bar b}$, 
and $\Ht^\al_{a \bar b}$.
Both $\aL^\al_{ab}$ and $\gt^\ab_{a \bar b}$ 
also control CPT violation.
Taken alone,
the coefficients 
$\aL^\al_{ab}$ and $\Ht^\al_{a \bar b}$
generate oscillation phases 
proportional to $L$ but independent of $E$,
while $\cL^\ab_{ab}$ and $\gt^\ab_{a \bar b}$
produce phases proportional to the product $LE$.
This indicates that 
experiments with long baselines or high energies 
are of special interest for studies
of massive Lorentz-violating models. 
However,
the techniques outlined in this work apply for any baseline
for which the perturbative approximation is valid,
including ones where oscillations due to mass are negligible.
Indeed,
the expressions for oscillation probabilities presented here 
reduce to those obtained for massless Lorentz-violating models  
\cite{km3}
in the limit of vanishing mass mixing.

Massive Lorentz-violating models 
can exhibit effects lying in any of the six classes 
of physical effects due to Lorentz and CPT violation
\cite{km1}.
All coefficients affect the spectral dependence
in at least some part of the energy range. 
Many of the associated operators violate rotation invariance,
which can produce direction-dependent oscillations.
For some experiments,
including ones with neutrino beams,
the daily rotation of the Earth
induces variations in time of the probabilities 
at multiples of the sidereal frequency. 
Both CPT violation
and neutrino-antineutrino mixings can occur.

In this work,
we show that for massive Lorentz-violating models
the coefficients 
$\aL^\al_{ab}$ and $\cL^\ab_{ab}$
primarily affect neutrino-neutrino 
and antineutrino-antineutrino mixings,
with $\aL^\al_{ab}$ controlling first-order differences
between the two mixings due to perturbative CPT violation.
Since the original introduction 
of these SME coefficients
\cite{ck},
a substantial theoretical literature has developed
concerning their implication for neutrino behavior 
in the context of massive Lorentz-violating models.
Many works restrict attention
to the special isotropic limits
with only $\aL^T_{ab}$ or $\cL^{TT}_{ab}$ nonzero and real
\cite{cg,bpww,bbm,dg,gl,dgms,dgpg,gghm,gktw,dges,rnknkm,xm,bgpg,ag},
and in some cases also to two flavors.
A few consider also anisotropic effects
\cite{km1,mp,bef,gl2,hmw,bb}.
Here,
we treat the general case,
allowing all components of $\aL^\al_{ab}$ and $\cL^\ab_{ab}$
to be nonzero.

In contrast,
the dominant effects from 
$\gt^\ab_{a \bar b}$ and $\Ht^\al_{a \bar b}$
in massive Lorentz-violating models arise only at second order.
They involve neutrino-antineutrino mixing
and also nonconservation of lepton number.
A single flavor can therefore suffice
to produce oscillations. 
Indeed,
a simple analytical form is known 
for the mixing probability 
for the general one-flavor case including mass
\cite{km1}.
A few theoretical works have considered special 
massive Lorentz-violating models of this type
\cite{scsk,hmp}.
At present,
there are no published experimental constraints
on any of the coefficients
$\gt^\ab_{a \bar b}$ and $\Ht^\al_{a \bar b}$.
In this work,
we investigate the general case
and identify some potential signals for experimental searches. 

Overall,
most massive Lorentz-violating models remain viable.
Only a few percent of the available coefficient space
has been explored experimentally
\cite{lsndlv,sklv,minoslv}.
The methods described in the present work
demonstrate that access to essentially 
the whole coefficient space is available
via a combination of existing and future experiments.

This paper is organized as follows.
The basic theory and notation is presented 
in Sec.\ \ref{sec theory}.
Section \ref{sec ham} reviews the properties 
of the hamiltonian governing Lorentz and CPT violation
in neutrino oscillations. 
The perturbation series for the transition amplitude
is derived in Sec.\ \ref{sec pert},
while the resulting oscillation probabilities
are obtained in Sec.\ \ref{sec prob}.
Section \ref{sec ac} considers
first-order effects involving the coefficients
$\aL^\al$ and $\cL^\ab$.
The directional and sidereal dependences of the probabilities
are discussed in Sec.\ \ref{sec osc}.
Examples are provided for the case of three generations 
and its two-generation limit 
in Sec.\ \ref{sec 3nu}.
Asymmetries characterizing violations 
of the discrete symmetries CP and CPT 
are discussed in Sec.\ \ref{sec cpt}.
Section \ref{sec gH} investigates 
the second-order effects involving the coefficients
$\gt^\ab$ and $\Ht^\al$.
Oscillations with lepton-number violation
are studied in Sec.\ \ref{sec lnv},
while others are considered in Sec.\ \ref{sec lnc}.
Section \ref{sec sum} concludes with a summary.

\section{Basic theory}
\label{sec theory}

This section begins with a brief review 
of the description of Lorentz and CPT violation 
in neutrino oscillations,
assuming three generations of left-handed neutrinos
and their antineutrinos.
We then use time-dependent perturbation theory
to derive expressions 
for the transition amplitudes and oscillation probabilities
valid for small Lorentz and CPT violation.

\subsection{Hamiltonian}
\label{sec ham}

Violations of Lorentz and CPT invariance 
in oscillations of left-handed neutrinos and their antineutrinos
can be characterized 
by a 6$\times$6 effective hamiltonian $(h_{\rm eff})_{AB}$ 
taking the form
\cite{km1}
\beq
(h_{\rm eff})_{AB}=(h_0)_{AB}+\de h_{AB}.
\label{ham}
\eeq
Here,
$h_0$ describes conventional Lorentz-invariant neutrino oscillations, 
while $\de h$ includes the Lorentz-violating contributions.
The uppercase indices take six values, 
$A,B,\ldots = e,\mu,\ta,\bar e,\bar \mu,\bar \ta$,
spanning the three flavors of neutrinos and antineutrinos.

Under typical assumptions,
the conventional term $h_0$ induces no mixing between
neutrinos and antineutrinos.
It is therefore block diagonal,
and we write it as
\beq
h_0=\left(\begin{array}{cc}
(h_0)_{ab} & 0 \\
0 & (h_0)_{\bar a\bar b}
\end{array}\right) 
=\frac{1}{2E}\left(\begin{array}{cc}
\De m^2_{ab} & 0 \\
0 & \De m^2_{\bar a\bar b}\\
\end{array}\right) ,
\label{h0}
\eeq
where $E$ is the neutrino energy,
lowercase indices $a,b,\ldots=e,\mu,\ta$ indicate neutrinos,
and lowercase barred indices
$\bar a,\bar b,\ldots=\bar e,\bar \mu,\bar \ta$ 
indicate antineutrinos.
The two 3$\times$3 mass matrices are related by
\beq
\De m^2_{\bar a\bar b}=\De m^{2\, *}_{ab},
\eeq
as required by the CPT theorem
\cite{owg}.
Note that contributions to the hamiltonian proportional to 
the unit matrix generate no oscillation effects,
but they may nonetheless be relevant 
to stability and causality of the underlying theory
\cite{kl}. 

The Lorentz-invariant hamiltonian $h_0$ 
can be diagonalized using a 6$\times$6 unitary matrix $U$,
\beq
(h_0)_{A'B'} =  \sum_{AB} U_{A'A} U^*_{B'B} (h_0)_{AB} ,
\eeq
where primed indices indicate the diagonal mass basis.
The absence of mixing between neutrinos and antineutrinos 
implies the mixing matrix is block diagonal,
\beq
U=\left(\begin{array}{cc}
U_{a'b} & 0\\
0 & U_{\bar a'\bar b}
\end{array}\right),
\eeq
with vanishing 3$\times$3 off-diagonal blocks,
\beq
U_{a' \bar b} = U_{\bar a' b} = 0.
\eeq
Since the mass matrices for neutrinos and antineutrinos 
are related by complex conjugation,
we also have 
\beq
U_{\bar a'\bar b}=U^*_{a'b}.
\eeq
The diagonal $3\times3$ blocks of $h_0$ can therefore be written as
\beq
(h_0)_{ab}=
(h_0)^*_{\bar a\bar b}=\sum_{a'=1,2,3} U^*_{a'a}U_{a'b}E_{a'},
\eeq
where $E_{a'}$ are the usual three neutrino eigenenergies.
In what follows,
we assume that these three eigenenergies are nondegenerate.
Note that this implies there are three twofold degeneracies
in the full $6\times6$ hamiltonian $(h_0)_{AB}$.

The Lorentz-violating term $\de h$ in Eq.\ \rf{ham}
can be written in the form
\beq
\de h=\left(\begin{array}{cc}
\de h_{ab} & \de h_{a\bar b} \\
\de h_{\bar ab} & \de h_{\bar a\bar b}
\end{array}\right).
\label{dh}
\eeq
For Lorentz-violating operators of renormalizable dimension,
the upper-left diagonal block takes the form
\beq
\de h_{ab}= 
\fr{1}{E}\big[\aL^\al p_\al -\cL^{\ab} p_\al p_\be \big]_{ab} 
\label{hnn} 
\eeq
and leads to mixing between neutrinos,
where the neutrino energy-momentum 4-vector is denoted 
$p_\al=(E,-\vec p)\approx E(1,-\hat p)$ 
and $\aL^{\al}_{ab}$, $\cL^{\ab}_{ab}$
are complex coefficients for Lorentz violation
\cite{km1}.
Hermiticity implies
\beq
\de h_{ab} = \de h_{ba}^*,
\eeq
which imposes the conditions 
\bea
\aL^\al_{ab} &=& \aL^{\al\, *}_{ba},
\nonumber\\
\cL^{\al\be}_{ab} &=& \cL^{\al\be\, *}_{ba}.
\eea
Similarly,
the lower-right diagonal block of $\de h$ 
produces mixing between antineutrinos,
\bea
\de h_{\bar a \bar b}&=& 
\fr{1}{E}\big[\aR^\al p_\al 
-\cR^{\ab} p_\al p_\be \big]_{\bar a\bar b}
\nonumber \\
&=& \fr{1}{E}\big[-\aL^\al p_\al 
-\cL^{\ab} p_\al p_\be \big]^*_{ab} .
\label{haa}
\eea

The off-diagonal 3$\times$3 blocks of $\de h$,
$\de h_{a\bar b}$ and $\de h_{\bar b a}$,
lead to neutrino-antineutrino mixing,
an unconventional effect.
These blocks 
obey the hermiticity condition
\beq
\de h_{a\bar b} = \de h_{\bar b a}^*
\eeq
and can be written as 
\cite{km1}
\bea
\de h_{a\bar b} &=&
-i\sqrt2 (\ep_+)_\al \big[\gt^\ab p_\be -\Ht^\al \big]_{a\bar b}, 
\nonumber\\
\de h_{\bar ab} &=&
i\sqrt2 (\ep_+)^*_\al \big[\gt^\ab p_\be -\Ht^\al \big]_{\bar ab}
\nonumber \\
&=& 
i\sqrt2 (\ep_+)^*_\al \big[\gt^\ab p_\be +\Ht^\al \big]^*_{a\bar b}.
\label{han}
\eea
In these equations,
the complex coefficients for Lorentz violation
$\gt^\ab_{a\bar b}$ and $\Ht^\al_{a\bar b}$
obey the relations
\bea
\gt^\ab_{a\bar b} 
&=& \gt^\ab_{b\bar a} =  \gt^{\ab\, *}_{\bar ba} , 
\nonumber\\
\Ht^\al_{a\bar b} 
&=& -\Ht^\al_{b\bar a} =  \Ht^{\al\, *}_{\bar ba}.
\eea
The complex 4-vector $(\ep_+)_\al=(0,-\vec\ep_+)$
represents the helicity state.
Introducing 
the local beam direction $\hat e_r$
and other unit vectors associated 
with local spherical coordinates as
\bea
\hat e_r &=& (\sin\th\cos\ph,\sin\th\sin\ph,\cos\th),
\nonumber\\
\hat e_\th &=& (\cos\th\cos\ph,\cos\th\sin\ph,-\sin\th),
\nonumber\\
\hat e_\ph &=& (-\sin\ph,\cos\ph,0),
\eea
the 3-vector $\vec\ep_+$ can be expressed as 
\beq
\vec\ep_+=\fr{1}{\sqrt 2}(\hat e_\th+i\hat e_\ph) .
\label{epplus}
\eeq

The coefficients $\aL^\al_{ab}$ and $\Ht^\al_{a\bar b}$
have dimensions of mass,
and each taken alone leads to oscillation effects 
that are energy independent.
In contrast,
the coefficients $\cL^\ab_{ab}$ and $\gt^\ab_{a\bar b}$
are dimensionless,
so their effects naturally scale with energy.
Note, 
however,
that combinations of coefficients can produce 
involved energy dependences,
including mimicking conventional mass terms
via the Lorentz-violating seesaw mechanism
\cite{km1,km2,kkt}. 

The coefficients 
$\aL^\al_{ab}$ and $\gt^\ab_{a\bar b}$ 
control CPT-odd effects,
while 
$\cL^\ab_{ab}$ and $\Ht^\al_{a\bar b}$ govern CPT-even ones.
Consequently,
CPT symmetry holds when
$\aL^\al_{ab}$ and $\gt^\ab_{a\bar b}$ vanish,
and we find the oscillation probabilities obey
the relationship
\beq
P_{\nu_a\rightarrow\nu_b} = P_{\nub_b\rightarrow\nub_a}
\quad
{\rm (CPT~invariance).}
\eeq
The CP symmetry may nonetheless be violated,
so the relation
$P_{\nu_a\rightarrow\nu_b} = P_{\nub_a\rightarrow\nub_b}$
may fail.
Further discussion of CP and CPT tests  
is provided in Sec.\ \ref{sec cpt} below.

All the coefficients discussed here
are taken to be spacetime constants,
so that translational symmetry
and energy-momentum conservation hold.
If the Lorentz violation is spontaneous
\cite{ks},
which may be ubiquitous in effective field theories
\cite{baak},
then the coefficients can be understood
as expectation values of operators 
in the fundamental theory.
Under these circumstances,
requiring constancy of the coefficients 
is equivalent to disregarding soliton solutions
and massive or Nambu-Goldstone (NG) modes
\cite{ng}.
When gravity is included,
the NG modes can play the role of the graviton
\cite{cardinal},
the photon in Einstein-Maxwell theory 
\cite{bumblebee},
or various new spin-dependent
\cite{ahclt}
or spin-independent
\cite{kt}
forces.
The presence of gravity can also produce
additional Lorentz-violating effects
on neutrino oscillations
\cite{akgrav,mmp}.

When neutrinos propagate in matter,
the resulting forward scattering 
on electrons, protons, and neutrons
can affect neutrino oscillations
\cite{tkjp}.
In the rest frame of the matter,
this adds to the effective hamiltonian $(h_{\rm eff})_{AB}$
terms equivalent to CPT-odd coefficients 
given by 
\cite{km1}
\bea
(a_{L,\rm eff})^0_{ee}&=&G_F(2n_e-n_n)/\sqrt{2} ,
\nonumber \\
(a_{L,\rm eff})^0_{\mu\mu}&=&(a_{L,\rm eff})^0_{\ta\ta}=
-G_Fn_n/\sqrt{2} ,
\label{matter}
\eea
where $n_e$ and $n_n$ 
are the number densities of electrons and neutrons
in the matter
and $G_F$ is the Fermi coupling constant.
For example,
in neutrino-oscillation experiments with long baselines,
the propagation is over comparatively long distances
in the Earth's crust.
In this case,
the densities $n_e$ and $n_n$ 
can be taken equal and constant to a good approximation,
with 
$\sqrt{2}G_Fn_e \simeq 2.1\times 10^{-22}$ GeV 
$\simeq$ (940 km)$^{-1}$.
In the perturbative analysis of Lorentz violation that follows,
any matter effects can be taken as part 
of the unperturbed hamiltonian $(h_0)_{AB}$.
In situations where mass oscillations dominate,
the matter effects could alternatively be treated as perturbative
and included as part of the Lorentz-violating term $\de h_{AB}$.

\subsection{Perturbation series}
\label{sec pert}

In this subsection,
we use standard techniques of time-dependent perturbation theory
to derive a perturbative series for the transition amplitudes.
We treat the hamiltonian component $\de h$ 
describing Lorentz and CPT violation
as small compared to $1/L$.

The time-evolution operator $S(t)$ is written in the form
\bea
S(t) &\equiv& e^{-i\heff t} 
\nonumber\\
&=& \big(e^{-i\heff t}e^{i h_0 t}\big)S^{(0)}(t) 
\nonumber\\
&=& S^{(0)}(t) + S^{(1)}(t) + S^{(2)}(t) + \cdots ,
\eea
where $S^{(n)}$ is the $n$th-order
perturbation in $\de h$.
The conventional term is given by 
\beq
S^{(0)}=e^{-ih_0t}.
\eeq
The higher-order terms are obtained
using the integral relation
\bea
e^{-i\heff t}e^{i h_0 t}
&=& 1 + \int_0^t dt_1 (-i)\De h(t_1) \nonumber\\
&&\hspace*{-65pt}
+ \int_0^t dt_2 \int_0^{t_2} dt_1 (-i)\De h(t_1)(-i)\De h(t_2)
+ \cdots ,
\label{ints}
\eea
where
\beq
\De h(t)=e^{-ih_0 t} \de h e^{ih_0 t} .
\eeq

The integrals \rf{ints} can conveniently be performed 
in the mass-diagonal basis.
To second order in Lorentz-violating coefficients,
the results take the form
\bea
\S{0}{A'B'} &=& \de_{A'B'} \T{0}{A'}(t) , 
\nonumber\\
\S{1}{A'B'} &=& -it\de h_{A'B'} \T{1}{A'B'}(t) , 
\nonumber \\
\S{2}{A'B'} &=& 
-\half t^2\sum_{C'}\de h_{A'C'}\de h_{C'B'} \T{2}{A'B'C'}(t) .
\eea
All sums over flavor indices are written explicitly 
throughout this work.
The time dependence is contained in the factors
$\T{0}{A'}$,
$\T{1}{A'B'}$,
and $\T{2}{A'B'C'}$.
The zeroth-order factor is the usual expression
\beq
\T{0}{A'}(t)=\exp(-iE_{A'}t).
\eeq
The first-order term is given by
\renewcommand{\arraystretch}{2.0}
\bea
\T{1}{A'B'}(t) &=&
\fr{1}{t} \exp(-iE_{B'}t)
\int_0^t dt_1 \exp(-i\De_{A'B'}t_1)
\nonumber \\
&& \hspace*{-40pt} =
\left\{\begin{array}{ll}
\exp(-iE_{B'}t), & E_{A'}= E_{B'},\\
\fr{\exp(-iE_{A'}t)-\exp(-iE_{B'}t)}{-i\De_{A'B'}t},
\quad & \mbox{otherwise,}
\end{array}\right.
\eea
where 
\beq
\De_{A'B'}=E_{A'}-E_{B'}
\eeq 
are the standard eigenenergy differences.
The second-order factor is given by the integral
\bea
\T{2}{A'B'C'}(t) &=& 
\fr{2}{t^2} \exp(-iE_{B'}t) 
\nonumber \\
&&\hspace*{-40pt}\times
\int_0^t dt_2 \int_0^{t_2} dt_1
\exp(-i\De_{A'C'}t_1)
\exp(-i\De_{C'B'}t_2)
\nonumber \\
&&\hspace*{-40pt}=
\left\{\begin{array}{ll}
\exp(-iE_{B'}t), & \hskip - 20pt E_{A'}=E_{B'}= E_{C'},\\
2\fr{\T{1}{A'B'}-\T{1}{C'B'}}{-i\De_{A'C'}t}
= 2\fr{\T{1}{A'C'}-\T{1}{A'B'}}{-i\De_{C'B'}t}, & \mbox{otherwise.}
\end{array}\right.
\nonumber \\
\eea
We give two expressions in the last case
so that expressions for the limiting cases
$E_{A'}=E_{C'}$ and $E_{B'}=E_{C'}$
can readily be extracted.
Note that both $\T{1}{A'B'}$ and $\T{2}{A'B'C'}$
are dimensionless functions of $E_{A'}t$
that are totally symmetric in mass-basis indices.
\renewcommand{\arraystretch}{1.0}

Transforming to the flavor basis,
the Lorentz-invariant transition amplitude is 
found to be
\beq
\S{0}{AB}=\sum_{A'} U^*_{A'A}U_{A'B} e^{-iE_{A'}t} .
\eeq
This leads to the usual oscillation probabilities 
for the Lorentz-invariant case of massive neutrinos.
At first order,
we choose to express the
transition amplitude in the convenient form
\bea
\S{1}{AB}(t) &\equiv & -it\H{1}{AB}(t) 
\nonumber\\
&=& -it\sum_{CD} \M{1}{AB}{CD} \de h_{CD} ,
\label{sone}
\eea
where the factors
\bea
\M{1}{AB}{CD}(t) &=&
\frac{1}{t}\int_0^{t_1}dt_1
\S{0}{AC}(t_1)\S{0}{DB}(t-t_1)
\label{mone}\\
&=& \sum_{A'B'}\T{1}{A'B'}(t)
U^*_{A'A} U_{A'C} U^*_{B'D} U_{B'B} 
\nonumber
\eea
depend on the energy and baseline of the experiment
and also on the conventional masses and mixing angles.
For given mass spectrum and mixing angles,
these factors determine the sensitivity of an experiment.
They are independent of the direction of the neutrino propagation
and of the coefficients for Lorentz violation.
As a result,
they remain unchanged as the Earth rotates.
The quantity $\H{1}{AB}$ defined in Eq.\ \rf{sone}
is a linear combination of these factors
and the Lorentz-violating perturbation $\de h$.
It plays a key role in the expressions 
for the oscillation probabilities
derived in the next subsection.
Note that 
$\H{1}{AB}$ reduces to $\de h_{AB}$
in the limit of negligible mass mixing.

The second-order result for the transition amplitude
can be written in a similar form.
We define
\bea
\S{2}{AB}(t) &\equiv& 
-\half t^2\H{2}{AB} 
\nonumber\\
&=&
-\half t^2\sum_{CDEF} \M{2}{AB}{CDEF} \de h_{CD} \de h_{EF} ,
\label{stwo}
\eea
where the experiment-dependent factors 
\bea
&&\hspace*{-20pt}
\M{2}{AB}{CDEF}(t) 
\nonumber\\
&=&
\frac{2}{t^2}\int^t_0 dt_2 \int^{t_2}_0 dt_1
\S{0}{AC}(t_1)\S{0}{DE}(t_2-t_1)\S{0}{FB}(t-t_2)
\nonumber\\
&=& \!\!\!\! \sum_{A'B'C'} \T{2}{A'B'C'}(t) \,
U^*_{A'A} U_{A'C} U^*_{C'D} U_{C'E} U^*_{B'F} U_{B'B} 
\nonumber\\
\eea
again determine the combinations of coefficients
relevant for oscillation effects.
In this case,
however,
the quantity $\H{2}{AB}$ defined in Eq.\ \rf{stwo}
is a quadratic combination of coefficients.
This leads to sidereal variations in neutrino oscillations 
at higher multiples of the Earth's rotation frequency.
Note that $\H{2}{AB}$ reduces to $(\de h^2)_{AB}$
in the limit of negligible mass mixing.

\subsection{Probabilities}
\label{sec prob}

Using the above results for the transition amplitudes,
we can derive the oscillation probabilities.
At zeroth order,
the transition amplitudes are Lorentz invariant
and take the usual block-diagonal form,
$\S{0}{a\bar b}=\S{0}{\bar ab}=0$.
So the zeroth-order probabilities 
for neutrino-antineutrino oscillations vanish,
\beq
P^{(0)}_{\nub_b\to\nu_a}=P^{(0)}_{\nu_b\to\nub_a} = 0 .
\label{zero1}
\eeq
Since CPT is conserved whenever Lorentz symmetry holds
\cite{owg},
we have
$\S{0}{ab}=\S{0}{\bar b \bar a}$.
This implies 
\beq
P^{(0)}_{\nu_b\to\nu_a}=P^{(0)}_{\nub_a\to\nub_b} = |\S{0}{ab}|^2 ,
\label{zero2}
\eeq
which leads to the usual results
for Lorentz-invariant oscillation probabilities 
in terms of mass-squared differences and mixing angles.

The full mixing probability is given by 
\beq
P_{\nu_B\to\nu_A}=|\S{0}{AB} + \S{1}{AB} + \S{2}{AB} + \cdots |^2.
\eeq
At second order in $\de h$,
this gives 
\bea
P^{(0)}_{\nu_B\to\nu_A} &=& |\S{0}{AB}|^2  ,
\nonumber \\
P^{(1)}_{\nu_B\to\nu_A} &=& 2\Re \big((\S{0}{AB})^* \S{1}{AB}\big)
\nonumber \\
&=&2t\, \Im \big((\S{0}{AB})^* \H{1}{AB}\big) , 
\nonumber \\
P^{(2)}_{\nu_B\to\nu_A} 
&=& 2\Re \big((\S{0}{AB})^* \S{2}{AB}\big) + |\S{1}{AB}|^2 ,
\nonumber \\
&=& -t^2\, \Re \big((\S{0}{AB})^* \H{2}{AB}\big) + t^2|\H{1}{AB}|^2 .
\label{sixdosc}
\eea
These equations involve the 6-dimensional space spanned by $A$.
They can be decomposed into oscillation probabilities 
expressed in terms of the neutrino and antineutrino subspaces 
spanned by $a$ and $\bar a$.
 
At first order,
a short calculation shows that the
probabilities can be written 
\bea
P^{(1)}_{\nu_b\to\nu_a} 
&=& 2t\, \Im \big((\S{0}{ab})^* \H{1}{ab}\big) ,
\nonumber\\
P^{(1)}_{\nub_b\to\nub_a} 
&=& 2t\, \Im \big((\S{0}{\bar a\bar b})^* \H{1}{\bar a\bar b}\big) ,
\nonumber\\
P^{(1)}_{\nu_b\to\nub_a} &=& P^{(1)}_{\nub_b\to\nu_a} = 0 .
\label{P1na}
\eea
No neutrino-antineutrino mixing occurs 
because $\S{0}{a\bar b}=\S{0}{\bar ab}=0$.
Only the combinations 
$\H{1}{ab}$ and $\H{1}{\bar a\bar b}$
obtained from the definition \rf{sone}
contribute to these probabilities.
Explicitly,
we find
\bea
\H{1}{ab}=\sum_{cd} \M{1}{ab}{cd} \de h_{cd} ,
\nonumber\\
\H{1}{\bar a\bar b}=
\sum_{\bar c\bar d} \M{1}{\bar a\bar b}{\bar c\bar d} 
\de h_{\bar c\bar d} ,
\nonumber\\
\H{1}{\bar a b}=
\sum_{\bar c d} \M{1}{\bar a b}{\bar c d} \de h_{\bar c d} , 
\nonumber\\
\H{1}{a\bar b}
=\sum_{c\bar d} \M{1}{a\bar b}{c\bar d} \de h_{c\bar d} .
\label{H1na}
\eea
Although $\H{1}{\bar ab}$ and $\H{1}{a\bar b}$
are absent from the first-order probabilities,
we include their expressions here 
because they enter the second-order probabilities below.
Since the first-order results involve 
the diagonal blocks of $\de h$,
only the coefficients $\aL^\al_{ab}$ and $\cL^\ab_{ab}$
play a role.

Decomposing the results \rf{sixdosc}
reveals that the second-order probabilities are
\bea
P^{(2)}_{\nu_b\to\nu_a} &=& 
-t^2\,  \Re \big((\S{0}{ab})^* \H{2}{ab}\big)
+ t^2 |\H{1}{ab}|^2 ,
\nonumber\\
P^{(2)}_{\nub_b\to\nub_a} 
&=& -t^2\, \Re \big((\S{0}{\bar a\bar b})^* \H{2}{\bar a\bar b}\big)
+ t^2 |\H{1}{\bar a\bar b}|^2 ,
\nonumber\\
P^{(2)}_{\nu_b\to\nub_a} &=& t^2 |\H{1}{\bar ab}|^2 ,
\nonumber\\
P^{(2)}_{\nub_b\to\nu_a} &=& t^2 |\H{1}{a\bar b}|^2 ,
\label{P2an}
\eea
where
$\S{0}{a\bar b}=\S{0}{\bar ab}=0$
is used to simplify the last two.
The probabilities 
$P^{(2)}_{\nu_b\to\nu_a}$ and 
$P^{(2)}_{\nub_b\to\nub_a}$ 
include both leading-order contributions 
from $\H{2}{ab}$ and $\H{2}{\bar a\bar b}$
as well as higher-order contributions
from the combinations
$\H{1}{ab}$ and $\H{1}{\bar a\bar b}$.
Also,
nonzero mixing between neutrinos and antineutrinos appears,
giving sensitivity to the linear combinations 
$\H{1}{\bar ab}$ and $\H{1}{a\bar b}$.
This shows that the dominant effects
of the coefficients $\gt^\ab_{a\bar b}$ and $\Ht^\al_{a\bar b}$
appear only at second order.
Moreover,
$\aL^\al_{ab}$ or $\cL^\ab_{ab}$
play no role in neutrino-antineutrino mixing
at this order. 

Explicit expressions for 
$\H{2}{ab}$ and $\H{2}{\bar a\bar b}$
can be obtained by decomposing
the quadratic combinations $\H{2}{AB}$ 
defined in Eq.\ \rf{stwo}.
The structure of the factors $\M{2}{AB}{CDEF}$
and the form of the mixing matrix $U$
reduce the number of terms that contribute.
In particular,
we find that $\M{2}{AB}{CDEF}$ vanishes
unless the index pairs $\{AC\}$, $\{BF\}$, and $\{DE\}$  
lie in the same subspace.
This leads to 
\bea
\H{2}{ab}&=&
\sum_{cdef} \M{2}{ab}{cdef} \de h_{cd} \de h_{ef}
\nonumber \\
&&
+\sum_{c\bar d\bar ef} 
\M{2}{ab}{c\bar d\bar ef} \de h_{c\bar d} \de h_{\bar ef} ,
\nonumber \\
\H{2}{\bar a\bar b}&=&
\sum_{\bar c\bar d\bar e\bar f}
\M{2}{\bar a\bar b}{\bar c\bar d\bar e\bar f}
\de h_{\bar c\bar d} \de h_{\bar e\bar f}
\nonumber \\
&&
+\sum_{\bar c d e\bar f} \M{2}{\bar a\bar b}{\bar c d e\bar f}
\de h_{\bar c d} \de h_{ e\bar f} ,
\nonumber \\
\H{2}{\bar a b}&=&
\sum_{\bar c d e f} \M{2}{\bar a b}{\bar c d e f}
\de h_{\bar c d} \de h_{e f}
\nonumber \\
&&
+\sum_{\bar c \bar d \bar e f} \M{2}{\bar a b}{\bar c\bar d\bar e f}
\de h_{\bar c\bar d} \de h_{\bar e f} ,
\nonumber \\
\H{2}{a \bar b}&=&
\sum_{c d e\bar f} \M{2}{a\bar b}{c d e\bar f}
\de h_{c d} \de h_{e\bar f}
\nonumber \\
&&
+\sum_{c \bar d \bar e\bar f} \M{2}{a\bar b}{c\bar d\bar e\bar f}
\de h_{c\bar d} \de h_{\bar e\bar f} .
\label{H2na}
\eea
Note that
$\H{2}{\bar a b}$
and 
$\H{2}{a \bar b}$
are absent from the second-order probabilities
but are included here for completeness.
This implies that no cross terms between
$\de h_{ab}$,
$\de h_{\bar a\bar b}$, and
$\de h_{a\bar b}= \de h_{\bar b a}^*$
appear at second order.
In particular,
all appearances of the coefficients 
$\gt^\ab_{a\bar b}$ and $\Ht^\al_{a\bar b}$
arise as squares or as quadratic products with each other,
without accompanying factors of 
$\aL^\al_{ab}$ or $\cL^\ab_{ab}$.

\section{Coefficients $\aL^\al_{ab}$ and $\cL^\ab_{ab}$}
\label{sec ac}

In this section,
we consider effects originating from the coefficients
$\aL^\al_{ab}$ and $\cL^\ab_{ab}$.
These contribute only to neutrino-neutrino mixing and
to antineutrino-antineutrino mixing.
We focus here on the dominant signals,
which arise from the first-order oscillation probabilities
$P^{(1)}_{\nu_b\to\nu_a}$
and $P^{(1)}_{\nub_b\to\nub_a}$
given in Eq.\ \rf{P1na}.

The theoretical analysis presented in the previous section
applies to any scenario involving neutrino propagation.
However,
the key experimental signals are different 
for beam experiments,
solar-neutrino studies,
and cosmological observations.
For definiteness in this work,
we provide results in the context of beam experiments.

Section \ref{sec osc} establishes the key expressions 
describing sidereal variations
in the oscillation probabilities.
Illustrative examples involving all three generations
are provided in Sec.\ \ref{sec 3nu},
while the two-flavor case is considered in Sec.\ \ref{sec 2nu}.
The construction of CP and CPT asymmetries
to characterize the effects is
considered in Sec.\ \ref{sec cpt}.

\subsection{Sidereal variations}
\label{sec osc}

The combinations of coefficients for Lorentz and CPT violation
entering the nonzero first-order probabilities \rf{P1na}
are controlled by the experimental factors
$\M{1}{ab}{cd}$ and
$\M{1}{\bar a\bar b}{\bar c\bar d}$
entering Eq.\ \rf{H1na}.
These factors can be calculated from Eq.\ \rf{mone}.
The time $t$ can be set equal to the baseline distance $L$ 
because any difference between the two is small 
and leads only to suppressed higher-order corrections.
For given values of $E$ and $L$,
the nine complex constants $\M{1}{ab}{cd}$
determine the coefficient combinations 
relevant for $\nu\mix\nu$ mixing,
while the nine complex constants
$\M{1}{\bar a\bar b}{\bar c\bar d}$
determine those for $\nub\mix\nub$ mixing.
If CP is conserved in the usual mass matrix,
as occurs in the two-generation limit,
then the mixing matrices are real
and obey $U_{ab}=U_{\bar a\bar b}$.
We then find 
$\M{1}{ab}{cd} = \M{1}{\bar a\bar b}{\bar c\bar d}$,
so a single set of nine constants determines
the experimentally relevant combinations
for both neutrinos and antineutrinos.

The specific combinations of 
$\aL^\al_{ab}$ and $\cL^\ab_{ab}$
appearing in the transition probabilities
can be found by considering the forms
of the relevant blocks of the hamiltonian,
Eqs.\ \rf{hnn} and \rf{haa}.
In this context,
it is convenient to define the linear combinations 
\bea
\AL^\al_{ab} = \sum_{cd} \M{1}{ab}{cd} \aL^\al_{cd} , 
\nonumber\\
\CL^\ab_{ab} = \sum_{cd} \M{1}{ab}{cd} \cL^\ab_{cd} , 
\nonumber\\
\AR^\al_{\bar a\bar b} = 
\sum_{\bar c\bar d}
\M{1}{\bar a\bar b}{\bar c\bar d} \aR^\al_{\bar c\bar d} , 
\nonumber\\
\CR^\ab_{\bar a\bar b} = 
\sum_{\bar c\bar d} 
\M{1}{\bar a\bar b}{\bar c\bar d} \cR^\ab_{\bar c\bar d} .
\eea
These are experiment-dependent
combinations of the fundamental coefficients
$\aL^\al_{ab}$ and $\cL^\ab_{ab}$.
Using $\AL^\al_{ab}$, $\AR^\al_{ab}$, $\CL^\ab_{ab}$,
and $\CR^\ab_{ab}$,
the combinations
$\H{1}{ab}$ and $\H{1}{\bar a\bar b}$
controlling the first-order probabilities
can be written in a form that mimics
the hamiltonian perturbations
$\de h_{ab}$ and $\de h_{\bar a\bar b}$,
\bea
\H{1}{ab}&=& 
\fr{1}{E}\big[\AL^\al p_\al -\CL^{\ab} p_\al p_\be \big]_{ab} , 
\nonumber
\\
\H{1}{\bar a \bar b}&=& 
\fr{1}{E}\big[\AR^\al p_\al 
-\CR^{\ab} p_\al p_\be \big]_{\bar a\bar b} .
\label{formone}
\eea
This form reveals the explicit 4-momentum dependence
of the transition probabilities. 

The momentum dependence implies that the mixing behavior 
can depend on the direction of neutrino propagation.
For Earth-based experiments,
the source and detector rotate at the sidereal frequency
$\om_\oplus\simeq 2\pi/(23$ h 56 min),
which can induce sidereal variations
in the oscillation probabilities.
Since the first-order probabilities are linear 
in $\H{1}{ab}$ or $\H{1}{\bar a \bar b}$,
each of which has both linear and quadratic terms 
in the 3-momentum,
sidereal variations controlled by the coefficients 
$\AL^\al_{ab}$ and $\CL^\ab_{ab}$
can occur at the frequencies
$\om_\oplus$ and $2\om_\oplus$.

To display explicitly these variations,
a choice of inertial frame must be specified.
By convention and convenience,
the standard inertial frame 
is taken as a Sun-centered celestial-equatorial frame
with coordinates $(T,X,Y,Z)$
\cite{tables,sunframe}.
The $Z$ axis of this frame is directed north and parallel
to the rotational axis of the Earth.
The $X$ axis points from the Sun towards the vernal equinox,
while the $Y$ axis completes a right-handed system.
The origin of the time coordinate 
is chosen as the vernal equinox 2000.
The Earth's rotation causes the neutrino 3-momentum 
to vary in local sidereal time $T_\oplus$ 
at the frequency $\om_\oplus$
in the Sun-centered frame,
unless it happens to lie along the $Z$ axis.

For neutrino-neutrino mixing,
we can display explicitly the sidereal variation 
by expanding $\H{1}{ab}$ as 
\bea
\H{1}{ab} &=&  \Cn{1}{ab}
\nonumber \\
&&
+\Asn{1}{ab} \sin\om_\oplus T_\oplus
+\Acn{1}{ab} \cos\om_\oplus T_\oplus
\nonumber \\
&&
+\Bsn{1}{ab} \sin2\om_\oplus T_\oplus
+\Bcn{1}{ab} \cos2\om_\oplus T_\oplus .
\qquad
\label{H1sid}
\eea
Suppose the neutrinos of interest are emitted 
in a definite direction relative to the Earth,
perhaps as a neutrino beam from an accelerator.
Let the vector $(\Nh^X,\Nh^Y,\Nh^Z)$
represent the propagation direction 
in the Sun-centered frame at local sidereal time $T_\oplus =0$.
We can write this vector in terms
of local spherical coordinates at the detector.
Denote by $\ch$ the colatitude of the detector.
Introduce at the detector 
the angle $\th$ between the beam direction and vertical,
and also the angle $\ph$ between the beam and east of south.
The components of the vector can then be written as 
\bea
\Nh^X 
&=&
\cos\ch\sin\th\cos\ph+\sin\ch\cos\th ,
\nonumber\\
\Nh^Y 
&=&
\sin\th\sin\ph,
\nonumber\\
\Nh^Z
&=&
-\sin\ch\sin\th\cos\ph+\cos\ch\cos\th.
\label{nvector}
\eea
Using these expressions,
the amplitudes in the expansion \rf{H1sid}
are specified in terms of coefficients for Lorentz violation 
in the Sun-centered frame as
\bea
\Cn{1}{ab} &=&
\AL^T_{ab}
- \Nh^Z \AL^Z_{ab}
\nonumber \\ 
&&
- \half(3-\Nh^Z\Nh^Z) E \CL^{TT}_{ab}
+ 2\Nh^Z E \CL^{TZ}_{ab}
\nonumber \\
&&
+ \half(1-3\Nh^Z\Nh^Z) E \CL^{ZZ}_{ab} ,
\nonumber\\
\Asn{1}{ab} &=&
\Nh^Y \AL^X_{ab}
-\Nh^X \AL^Y_{ab}
\nonumber \\ 
&&
-2\Nh^Y E \CL^{TX}_{ab}
+2\Nh^X E \CL^{TY}_{ab}
\nonumber \\ 
&&
+2\Nh^Y \Nh^Z E \CL^{XZ}_{ab}
-2\Nh^X \Nh^Z E \CL^{YZ}_{ab} ,
\nonumber \\ 
\Acn{1}{ab} &=&
-\Nh^X \AL^X_{ab}
-\Nh^Y \AL^Y_{ab}
\nonumber \\ 
&&
+2\Nh^X E \CL^{TX}_{ab}
+2\Nh^Y E \CL^{TY}_{ab}
\nonumber \\ 
&&
-2\Nh^X \Nh^Z E \CL^{XZ}_{ab}
-2\Nh^Y \Nh^Z E \CL^{YZ}_{ab} ,
\nonumber \\ 
\Bsn{1}{ab} &=&
\Nh^X\Nh^Y E \big(\CL^{XX}_{ab}-\CL^{YY}_{ab}\big)
\nonumber \\ 
&&
-\big(\Nh^X\Nh^X-\Nh^Y\Nh^Y\big) E \CL^{XY}_{ab} ,
\nonumber \\ 
\Bcn{1}{ab} &=&
-2\Nh^X\Nh^Y E \CL^{XY}_{ab} 
\nonumber \\ 
&&
\hskip-10pt
-\half\big(\Nh^X\Nh^X-\Nh^Y\Nh^Y\big) E \big(\CL^{XX}_{ab}
-\CL^{YY}_{ab}\big) .
\nonumber\\
\label{H1Bc}
\eea
The form of the above expansion matches 
that used in Ref.\ \cite{km3}
in the context of short-baseline neutrino experiments.

The sidereal variation in 
$\H{1}{ab}$ leads
to a corresponding variation
in the probabilities.
We parametrize these variations as 
\bea
\frac{P^{(1)}_{\nu_b\to\nu_a}}{2L} &=& 
\Im \big((\S{0}{ab})^* \H{1}{ab}\big)
\nonumber \\ 
&&
\hskip-40pt
=\PCn{1}{ab}
+\PAsn{1}{ab} \sin\om_\oplus T_\oplus
+\PAcn{1}{ab} \cos\om_\oplus T_\oplus
\nonumber \\
&&
+\PBsn{1}{ab} \sin2\om_\oplus T_\oplus
+\PBcn{1}{ab} \cos2\om_\oplus T_\oplus  ,
\nonumber \\ 
\label{P1sid} 
\eea
where
\bea
\PCn{1}{ab} &=& \Im \big((\S{0}{ab})^* \Cn{1}{ab}\big) ,   
\nonumber \\ 
\PAsn{1}{ab} &=& \Im \big((\S{0}{ab})^* \Asn{1}{ab}\big) , 
\nonumber \\ 
\PAcn{1}{ab} &=& \Im \big((\S{0}{ab})^* \Acn{1}{ab}\big) , 
\nonumber \\ 
\PBsn{1}{ab} &=& \Im \big((\S{0}{ab})^* \Bsn{1}{ab}\big) , 
\nonumber \\ 
\PBcn{1}{ab} &=& \Im \big((\S{0}{ab})^* \Bcn{1}{ab}\big)  
\label{P1Bc}
\eea
are combination of coefficients for Lorentz violations.
Note that the sidereal amplitudes
$\PCn{1}{ab}$, $\PAsn{1}{ab}$, $\PAcn{1}{ab}$,
$\PBsn{1}{ab}$, and $\PBcn{1}{ab}$
are tiny,
with size determined by 
$\aL^\al_{ab}$ and $\cL^\ab_{ab}$.
The expression \rf{P1sid} 
reveals that the experimental sensitivity
to perturbative Lorentz and CPT violation
increases with the baseline $L$.

For antineutrino-antineutrino oscillations,
analogous results hold.
We can expand $\H{1}{\bar a \bar b}$ 
in the form \rf{H1sid},
replacing the indices $\{ab\}$ with $\{\bar a\bar b\}$.
The amplitudes again take the form \rf{H1Bc},
but with the substitutions 
$\AL \to \AR$, $\CL \to \CR$,
$\{ab\} \to \{\bar a\bar b\}$.
Similarly,
the sidereal variation in the probabilities
can be written as Eq.\ \rf{P1sid}
by replacement of the indices.

\subsection{Illustrations}
\label{sec 3nu}

In this subsection,
the first-order perturbative formalism derived above
is applied to several illustrative situations.
Following some comments about the methodology,
we consider an example involving mixing 
of all three flavors of neutrinos
and then discuss the limiting case of two flavors.

\subsubsection{Methodology}
\label{sec mass}

Since the effects from Lorentz and CPT violation
are perturbative,
an explicit analysis must 
specify the conventional mass spectrum and mixing angles.
In the standard three-neutrino massive model
\cite{pdg}, 
the usual $3\times3$ effective hamiltonian $(h_0)_{ab}$
for neutrino-neutrino vacuum mixing appears 
as the upper-left block of Eq.\ \rf{h0}.
It can be written as
\bea
(h_0)_{ab} &=& \fr{1}{2E} \De m^2_{ab}
\nonumber \\
&=&\fr{1}{2E} \sum_{a'b'} U^*_{a'a} U_{b'b} \De m^2_{a'b'} ,
\label{HSM} 
\eea
where $\De m^2_{a'b'}$ is the diagonal mass matrix.
Only two mass-squared differences contribute
to oscillations.
Without loss of generality,
we can therefore express the diagonal mass matrix as
\beq
m^2_{a'b'}=
\begin{pmatrix} 0&0&0&\\
0 & \De m_{\odot}^2 & 0 \\
0 & 0 & \De m_{\text{atm}}^2
\end{pmatrix} .
\label{m_ab(3)}
\eeq
For a normal mass hierarchy,
the quantity $\De m_{\odot}^2$
is the smaller of the two mass-squared differences.
It is of particular relevance 
in situations involving low-energy oscillations
such as solar-neutrino measurements.
The larger mass-squared difference $\De m_{\text{atm}}^2$
plays a central role where mixing of high-energy neutrinos occurs,
such as atmospheric-neutrino experiments.
The mixing matrix $U_{a'b}$ can be written as
\bea
U_{a'b}&=& 
\begin{pmatrix} 
c_{12} & -s_{12} & 0 \\ s_{12} & c_{12} & 0\\ 0&0&1
\end{pmatrix}% 
\begin{pmatrix} 
c_{13}&0&-s_{13}e^{-i\de}\\0&1&0\\s_{13}e^{i\de}&0&c_{13}
\end{pmatrix}% 
\nonumber \\
&& \hspace*{40pt} \times
\begin{pmatrix} 
1&0&0 \\ 0& c_{23} & -s_{23} \\ 0& s_{23} & c_{23} 
\end{pmatrix} ,
\eea
where $c_{ij}=\cos\theta_{ij}$, 
$s_{ij}=\sin\theta_{ij}$, and $\de$
is the CP-violating phase.
For antineutrino-antineutrino mixing, 
the conventional effective hamiltonian for vacuum mixing
is obtained by complex conjugation of the above results,
which is equivalent to changing the sign of $\de$.

In the presence of matter,
the $3\times3$ effective hamiltonian $(h_0)_{ab}$ 
for neutrino-neutrino mixing 
acquires an additional term from Eq.\ \rf{matter}
and becomes 
\bea
(h_0)_{ab} &=& \fr{1}{2E} \De m^2_{ab}
+(a_{L,\rm eff})^T_{ee} \de_{ae}\de_{be} ,
\label{HSMmatter} 
\eea
where the index $e$ refers to the electron-neutrino flavor.
The additional term changes the eigenvalues of $(h_0)_{ab}$
and hence the explicit values of the components 
of the overall mixing matrix.
Note that the expression \rf{HSMmatter}
is strictly valid only in the rest frame of the matter.
For the specific scenarios considered below,
this frame comoves with the Earth as it rotates on its axis
and revolves about the Sun.
These motions are nonrelativistic,
however,
so the comoving frame can be identified 
with the Sun-centered frame of Sec.\ \ref{sec osc}
to an accuracy of parts in $10^4$.
We adopt this identification in what follows.

In performing an analysis for Lorentz and CPT violation,
the appropriate methodology depends on
the location of the experiment in $L$-$E$ space
and on the presently unknown value of $\th_{13}$.
Consider, for example,
three hypothetical beam experiments 
with the same long baseline $L \sim 200$ km
but with differing energies 
$E_1 \sim 10$ MeV,
$E_2 \sim 1$ GeV,
$E_3 \sim 100$ GeV.
The first lies in a region where
mass oscillations involve all three flavors,
so the treatment of Lorentz and CPT violation
requires the three-flavor formalism of the previous section.
The same is true of the second experiment
for large $\th_{13}$.
However,
if $\th_{13}$ is small or zero,
then significant mass mixings in this second experiment
involve only two generations.
Lorentz-violating effects within these two generations
can then be studied using a two-flavor limit 
of the previous section,
while effects involving the third flavor
are well described using the procedures for negligible mass mixing 
presented in Ref.\ \cite{km3}.
For the third experiment,
no significant mass mixings occur
and so the methods of Ref.\ \cite{km3}
are applicable for Lorentz and CPT violation 
in all flavors of oscillations.

High sensitivity to operators for Lorentz and CPT violation
of mass dimension three,
such as those controlled by the coefficients $\aL^\al_{ab}$,
can be achieved in experiments with long baselines $L$.
Similarly,
high sensitivity to operators of mass dimension four,
such as those governed by $\cL^\ab_{ab}$,
can be obtained via long baselines $L$,
high energies $E$, or both.
At present,
most existing or planned long-baseline experiments 
have baselines $L\sim$ 200-1500 km
and energies $E\sim$ 1-10 GeV
and hence lie in a region of $L$-$E$ space
analogous to that of the second hypothetical experiment above.

For illustrative purposes,
we consider here a variety of beam experiments
involving long baselines $L$
and seeking $\nu_e$ appearance in $\nu_\mu$ beams
or studying $\nu_\mu$ disappearance. 
Existing experiments in this category include
KEK to Super-Kamiokande (K2K)
with baseline $L\simeq 250$ km 
\cite{K2K},
the MINOS far detector
with baseline $L\simeq 750$ km 
\cite{MINOS}, 
and the Oscillation Project with Emulsion-Tracking Apparatus
(OPERA)
\cite{OPERA}.
The latter has essentially identical baseline $L\simeq 750$ km 
to the Imaging Cosmic and Rare Underground Signals
experiment (ICARUS)
\cite{ICARUS}.
The Fermilab E929 experiment (NO$\nu A$)
\cite{NOVA}
with baseline $L\simeq 800$ km 
is currently under construction,
while the Tokai to Kamioka (T2K) experiment
\cite{T2K}
with baseline $L\simeq 300$ km 
has recently begun data taking.
Other experiments with even longer baselines
are under consideration,
including
one at the Deep Underground Science and Engineering Lab (DUSEL)
\cite{DUSEL}
using a neutrino beam from Fermilab
and baseline $L\sim1300$ km,
and the Tokai to Kamioka and Korea (T2KK) experiment
\cite{T2KK}
using the same neutrino beam as T2K 
but with baseline $L\simeq 1000$ km. 
All of these experiments 
have excellent sensitivity to
perturbative Lorentz and CPT violation.

For definiteness in what follows,
we consider two explicit scenarios.
In the first,
discussed in Sec.\ \ref{sec app},
we take a comparatively large value of $\th_{13}$
and consider the effects of Lorentz and CPT violation
for studies of $\nu_e$ appearance.
For this situation,
mass mixing involving all three flavors occurs
and so the full formalism 
of the previous section is appropriate for the analysis. 
In the second scenario,
considered in Sec.\ \ref{sec 2nu},
we suppose $\th_{13}$ is negligible
and investigate the effects of Lorentz and CPT violation
on $\nu_\mu$ disappearance.
For this case,
only $\nu_\mu\mix \nu_\ta$ involves significant mass mixing 
and so a two-flavor limit of the previous section
can be applied.

We can also identify two interesting regions of $L$-$E$ space 
that could benefit from the development of new experiments.
The first is the region of long baselines $L\gsim$ 200 km
with low energies $E\lsim$ 100 MeV.
This is of particular interest if $\th_{13} \simeq 0^\circ$,
since it provides an opportunity 
for clean studies of three-flavor mixings
that are otherwise challenging to perform.
A comparatively intense source is needed  
due to the long baseline 
and the cross-section falloff with energy.
One possibility is a setup similar to the
Kamioka Liquid Scintillator Antineutrino Detector (KamLAND)
\cite{kamland}
but with a single source, 
directional sensitivity,
or both. 
For longer baselines,
a beta beam 
\cite{beta}
may be an interesting option.
A low-energy beta beam has been studied 
in the context of short-baseline experiments
\cite{cv}.
Another possibility might be 
an intense pulsed neutrino beam such as that proposed for 
a neutrino facility at the Spallation Neutron Source ($\nu$-SNS)
\cite{sns}
and lying in the 10-50 MeV energy range. 

The second interesting region lies 
at high energies $E\gsim 100$ GeV,
even for comparatively short baselines $L\lsim$ 10 km.
Neutrino beams at these energies 
have been used in the Neutrinos at the Tevatron (NuTeV)
\cite{nutev}
and Chicago-Columbia-Fermilab-Rochester (CCFR) 
\cite{ccfr}
experiments.
For studies of the coefficients $\cL^\ab_{ab}$ 
for Lorentz and CPT violation,
a high energy compensates for a shorter baseline
and so oscillation experiments in this region
could have a competitive reach.
Since mass mixing is negligible,
the methods of Ref.\ \cite{km3} are applicable
for analyzing Lorentz and CPT violation in this case.

\begin{table*}
\begin{center}
\begin{tabular}{c|ccccccc}
Experiment & K2K & MINOS & OPERA, & NO$\nu$A & T2K & DUSEL & T2KK \\
& & & ICARUS & & & & \\
\hline
$\Re \M{1}{e\mu}{ee}$           &   -0.05   &   -0.10   &   -0.01   &   -0.17   &   -0.16   &   -0.08   &   -0.11   \\
$\Re \M{1}{e\mu}{e\mu}$             &   0.84    &   0.63    &   0.88    &   0.38    &   0.48    &   0.38    &   0.41    \\
$\Re \M{1}{e\mu}{e\tau}$            &   -0.13   &   -0.24   &   -0.02   &   -0.46   &   -0.46   &   -0.35   &   -0.39   \\
$\Re \M{1}{e\mu}{\mu e}$            &   0.00    &   0.00    &   0.00    &   0.00    &   0.00    &   0.01    &   0.01    \\
$\Re \M{1}{e\mu}{\mu \mu}$          &   -0.05   &   -0.09   &   -0.01   &   -0.15   &   -0.15   &   -0.09   &   -0.11   \\
$\Re \M{1}{e\mu}{\mu \tau}$         &   -0.01   &   -0.02   &   0.00    &   -0.01   &   -0.01   &   0.04    &   0.03    \\
$\Re \M{1}{e\mu}{\tau e}$           &   0.00    &   0.00    &   0.00    &   0.00    &   0.00    &   0.00    &   0.00    \\
$\Re \M{1}{e\mu}{\tau\mu}$          &   -0.05   &   -0.08   &   -0.01   &   -0.14   &   -0.14   &   -0.06   &   -0.08   \\
$\Re \M{1}{e\mu}{\tau\tau}$         &   -0.01   &   -0.01   &   0.00    &   0.00    &   0.00    &   0.03    &   0.02    \\
                                    &       &       &       &       &       &       &       \\[-5pt]
$\Im \M{1}{e\mu}{ee}$           &   -0.08   &   -0.06   &   -0.02   &   -0.03   &   -0.08   &   0.10    &   0.07    \\
$\Im \M{1}{e\mu}{e\mu}$             &   -0.39   &   -0.62   &   -0.44   &   -0.61   &   -0.43   &   -0.34   &   -0.32   \\
$\Im \M{1}{e\mu}{e\tau}$            &   -0.24   &   -0.24   &   -0.06   &   -0.21   &   -0.27   &   0.17    &   0.12    \\
$\Im \M{1}{e\mu}{\mu e}$            &   0.00    &   0.01    &   0.00    &   0.02    &   0.03    &   -0.04   &   -0.03   \\
$\Im \M{1}{e\mu}{\mu \mu}$          &   -0.07   &   -0.06   &   -0.02   &   -0.02   &   -0.02   &   -0.05   &   -0.05   \\
$\Im \M{1}{e\mu}{\mu \tau}$         &   0.01    &   0.02    &   0.00    &   0.07    &   0.08    &   -0.08   &   -0.06   \\
$\Im \M{1}{e\mu}{\tau e}$           &   0.00    &   0.01    &   0.00    &   0.02    &   0.02    &   -0.05   &   -0.04   \\
$\Im \M{1}{e\mu}{\tau\mu}$          &   -0.06   &   -0.05   &   -0.02   &   0.00    &   0.00    &   -0.02   &   -0.02   \\
$\Im \M{1}{e\mu}{\tau\tau}$         &   0.01    &   0.02    &   0.00    &   0.06    &   0.07    &   -0.12   &   -0.10   \\ 
\end{tabular}
\caption{\label{Table A}
Approximate numerical values 
of experiment-dependent dimensionless factors
$\M{1}{e\mu}{c d}$
for the experiments 
K2K, MINOS, OPERA, ICARUS, NO$\nu$A, T2K, DUSEL, and T2KK.
Numerical values are computed via Eq.\ \rf{mone}
adopting the parameters \rf{theta23=45},
using estimated beam baselines $L$
and neutrino energies $E$ for each experiment.
Within this approximation,
the antineutrino factors
$\M{1}{\bar e\bar \mu}{\bar c\bar d}$
are identical.} 
\end{center} 
\end{table*}

\begin{table*}
\begin{center}
\begin{tabular}{c|cccc|cccc|cccc}
Experiment&\multicolumn{4}{c|}{K2K} & \multicolumn{4}{c|}{MINOS} &\multicolumn{4}{c}{OPERA, ICARUS} \\
\hline
Amplitude&
$\PASN{1}$ & $\PACN{1}$ & $\PBSN{1}$ & $\PBCN{1}$ &
$\PASN{1}$ & $\PACN{1}$ & $\PBSN{1}$ & $\PBCN{1}$ &
$\PASN{1}$ & $\PACN{1}$ & $\PBSN{1}$ & $\PBCN{1}$ \\ &&&&&&&&&&&\\[-10pt]
\hline
$\Re\AL^X_{e\mu}$	&	$	-0.1	$&$	0.0	$&$	 - 	$&$	 - 	$&$	0.0	$&$	0.1	$&$	 - 	$&$	 - 	$&$	0.0	$&$	0.0	$&$	 - 	$&$	 - 	$	\\
$\Re\AL^Y_{e\mu}$	&	$	0.0	$&$	0.1	$&$	 - 	$&$	 - 	$&$	0.1	$&$	0.0	$&$	 - 	$&$	 - 	$&$	0.0	$&$	0.0	$&$	 - 	$&$	 - 	$	\\
$\Re\CL^{TX}_{e\mu}$	&	$	0.4	$&$	0.0	$&$	 - 	$&$	 - 	$&$	0.2	$&$	-0.3	$&$	 - 	$&$	 - 	$&$	-1.0	$&$	0.6	$&$	 - 	$&$	 - 	$	\\
$\Re\CL^{TY}_{e\mu}$	&	$	0.0	$&$	-0.4	$&$	 - 	$&$	 - 	$&$	-0.3	$&$	-0.2	$&$	 - 	$&$	 - 	$&$	0.6	$&$	1.0	$&$	 - 	$&$	 - 	$	\\
$\Re\CL^{XX}_{e\mu}$	&	$	- 	$&$	 - 	$&$	0.0	$&$	0.1	$&$	 - 	$&$	 - 	$&$	0.1	$&$	0.0	$&$	 - 	$&$	 - 	$&$	0.2	$&$	0.1	$	\\
$\Re\CL^{XY}_{e\mu}$	&	$	- 	$&$	 - 	$&$	0.0	$&$	-0.1	$&$	 - 	$&$	 - 	$&$	-0.1	$&$	0.0	$&$	 - 	$&$	 - 	$&$	-0.2	$&$	-0.1	$	\\
$\Re\CL^{XZ}_{e\mu}$	&	$	0.0	$&$	0.0	$&$	 - 	$&$	 - 	$&$	-0.1	$&$	0.2	$&$	 - 	$&$	 - 	$&$	-0.4	$&$	0.2	$&$	 - 	$&$	 - 	$	\\
$\Re\CL^{YY}_{e\mu}$	&	$	- 	$&$	 - 	$&$	0.2	$&$	0.0	$&$	 - 	$&$	 - 	$&$	0.0	$&$	-0.1	$&$	 - 	$&$	 - 	$&$	0.3	$&$	-0.4	$	\\
$\Re\CL^{YZ}_{e\mu}$	&	$	0.0	$&$	0.0	$&$	 - 	$&$	 - 	$&$	0.2	$&$	0.1	$&$	 - 	$&$	 - 	$&$	0.2	$&$	0.4	$&$	 - 	$&$	 - 	$	\\
&&&&&&&&&&&&	\\[-5pt]
$\Im\AL^X_{e\mu}$	&	$	0.1	$&$	0.0	$&$	 - 	$&$	 - 	$&$	0.1	$&$	-0.1	$&$	 - 	$&$	 - 	$&$	0.0	$&$	0.0	$&$	 - 	$&$	 - 	$	\\
$\Im\AL^Y_{e\mu}$	&	$	0.0	$&$	-0.1	$&$	 - 	$&$	 - 	$&$	-0.1	$&$	-0.1	$&$	 - 	$&$	 - 	$&$	0.0	$&$	0.0	$&$	 - 	$&$	 - 	$	\\
$\Im\CL^{TX}_{e\mu}$	&	$	-0.3	$&$	0.0	$&$	 - 	$&$	 - 	$&$	-0.4	$&$	0.5	$&$	 - 	$&$	 - 	$&$	0.5	$&$	-0.3	$&$	 - 	$&$	 - 	$	\\
$\Im\CL^{TY}_{e\mu}$	&	$	0.0	$&$	0.0	$&$	 - 	$&$	 - 	$&$	0.2	$&$	-0.3	$&$	 - 	$&$	 - 	$&$	0.2	$&$	-0.1	$&$	 - 	$&$	 - 	$	\\
$\Im\CL^{XX}_{e\mu}$	&	$	- 	$&$	 - 	$&$	0.0	$&$	-0.1	$&$	 - 	$&$	 - 	$&$	-0.1	$&$	0.0	$&$	 - 	$&$	 - 	$&$	-0.1	$&$	-0.1	$	\\
$\Im\CL^{XY}_{e\mu}$	&	$	- 	$&$	 - 	$&$	0.0	$&$	0.1	$&$	 - 	$&$	 - 	$&$	0.1	$&$	0.0	$&$	 - 	$&$	 - 	$&$	0.1	$&$	0.1	$	\\
$\Im\CL^{XZ}_{e\mu}$	&	$	0.0	$&$	0.0	$&$	 - 	$&$	 - 	$&$	0.2	$&$	-0.3	$&$	 - 	$&$	 - 	$&$	0.2	$&$	-0.1	$&$	 - 	$&$	 - 	$	\\
$\Im\CL^{YY}_{e\mu}$	&	$	- 	$&$	 - 	$&$	-0.1	$&$	0.0	$&$	 - 	$&$	 - 	$&$	0.1	$&$	0.2	$&$	 - 	$&$	 - 	$&$	-0.1	$&$	0.3	$	\\
$\Im\CL^{YZ}_{e\mu}$	&	$	0.0	$&$	0.0	$&$	 - 	$&$	 - 	$&$	-0.3	$&$	-0.2	$&$	 - 	$&$	 - 	$&$	-0.1	$&$	-0.2	$&$	 - 	$&$	 - 	$	\\
\hline
Sensitivity &\multicolumn{4}{c|}{$<8\times10^{-23}$} & \multicolumn{4}{c|}{$<3\times10^{-23}$}&\multicolumn{4}{c}{$<3\times10^{-23}$}  \\
\end{tabular} 
\vskip 20pt
\begin{tabular}{c|cccc|cccc|cccc|cccc}
Experiment&\multicolumn{4}{c|}{NO$\nu$A} & \multicolumn{4}{c|}{T2K} &\multicolumn{4}{c|}{DUSEL} & \multicolumn{4}{c}{T2KK} \\
\hline
Amplitude&
$\PASN{1}$ & $\PACN{1}$ & $\PBSN{1}$ & $\PBCN{1}$ &
$\PASN{1}$ & $\PACN{1}$ & $\PBSN{1}$ & $\PBCN{1}$ &
$\PASN{1}$ & $\PACN{1}$ & $\PBSN{1}$ & $\PBCN{1}$ &
$\PASN{1}$ & $\PACN{1}$ & $\PBSN{1}$ & $\PBCN{1}$ \\ &&&&&&&&&&&&&&&\\[-10pt]
\hline
$\Re\AL^X_{e\mu}$	&	$	0.0	$&$	0.0	$&$	 - 	$&$	 - 	$&$	0.0	$&$	0.0	$&$	 - 	$&$	 - 	$&$	-0.1	$&$	0.0	$&$	 - 	$&$	 - 	$&$	-0.1	$&$	0.0	$&$	 - 	$&$	 -  	$	\\
$\Re\AL^Y_{e\mu}$	&	$	0.0	$&$	0.0	$&$	 - 	$&$	 - 	$&$	0.0	$&$	0.0	$&$	 - 	$&$	 - 	$&$	0.0	$&$	0.1	$&$	 - 	$&$	 - 	$&$	0.0	$&$	0.1	$&$	 - 	$&$	 -  	$	\\
$\Re\CL^{TX}_{e\mu}$	&	$	0.0	$&$	0.0	$&$	 - 	$&$	 - 	$&$	0.0	$&$	0.0	$&$	 - 	$&$	 - 	$&$	0.1	$&$	0.0	$&$	 - 	$&$	 - 	$&$	0.1	$&$	0.0	$&$	 - 	$&$	 -  	$	\\
$\Re\CL^{TY}_{e\mu}$	&	$	0.0	$&$	0.0	$&$	 - 	$&$	 - 	$&$	0.0	$&$	0.0	$&$	 - 	$&$	 - 	$&$	0.0	$&$	-0.1	$&$	 - 	$&$	 - 	$&$	0.0	$&$	-0.1	$&$	 - 	$&$	 -  	$	\\
$\Re\CL^{XX}_{e\mu}$	&	$	 - 	$&$	 - 	$&$	0.0	$&$	0.0	$&$	 - 	$&$	 - 	$&$	0.0	$&$	0.0	$&$	 - 	$&$	 - 	$&$	0.0	$&$	0.0	$&$	 - 	$&$	 - 	$&$	0.0	$&$	0.0	$	\\
$\Re\CL^{XY}_{e\mu}$	&	$	 - 	$&$	 - 	$&$	0.0	$&$	0.0	$&$	 - 	$&$	 - 	$&$	0.0	$&$	0.0	$&$	 - 	$&$	 - 	$&$	0.0	$&$	0.0	$&$	 - 	$&$	 - 	$&$	0.0	$&$	0.0	$	\\
$\Re\CL^{XZ}_{e\mu}$	&	$	0.0	$&$	0.0	$&$	 - 	$&$	 - 	$&$	0.0	$&$	0.0	$&$	 - 	$&$	 - 	$&$	0.0	$&$	0.0	$&$	 - 	$&$	 - 	$&$	0.0	$&$	0.0	$&$	 - 	$&$	 -  	$	\\
$\Re\CL^{YY}_{e\mu}$	&	$	 - 	$&$	 - 	$&$	0.0	$&$	0.0	$&$	 - 	$&$	 - 	$&$	0.0	$&$	0.0	$&$	 - 	$&$	 - 	$&$	0.1	$&$	0.0	$&$	 - 	$&$	 - 	$&$	0.1	$&$	0.0	$	\\
$\Re\CL^{YZ}_{e\mu}$	&	$	0.0	$&$	0.0	$&$	 - 	$&$	 - 	$&$	0.0	$&$	0.0	$&$	 - 	$&$	 - 	$&$	0.0	$&$	0.0	$&$	 - 	$&$	 - 	$&$	0.0	$&$	0.0	$&$	 - 	$&$	 -  	$	\\
	&			&		&		&		&		&		&		&		&		&		&		&		&		&		&		&			\\[-5pt]
$\Im\AL^X_{e\mu}$	&	$	0.2	$&$	-0.2	$&$	 - 	$&$	 - 	$&$	0.3	$&$	0.0	$&$	 - 	$&$	 - 	$&$	0.1	$&$	0.0	$&$	 - 	$&$	 - 	$&$	0.2	$&$	0.0	$&$	 - 	$&$	 -  	$	\\
$\Im\AL^Y_{e\mu}$	&	$	-0.2	$&$	-0.2	$&$	 - 	$&$	 - 	$&$	0.0	$&$	-0.3	$&$	 - 	$&$	 - 	$&$	0.0	$&$	-0.1	$&$	 - 	$&$	 - 	$&$	0.0	$&$	-0.2	$&$	 - 	$&$	 -  	$	\\
$\Im\CL^{TX}_{e\mu}$	&	$	-0.6	$&$	0.8	$&$	 - 	$&$	 - 	$&$	-0.4	$&$	0.0	$&$	 - 	$&$	 - 	$&$	-0.3	$&$	0.0	$&$	 - 	$&$	 - 	$&$	-0.3	$&$	0.0	$&$	 - 	$&$	 -  	$	\\
$\Im\CL^{TY}_{e\mu}$	&	$	0.4	$&$	-0.5	$&$	 - 	$&$	 - 	$&$	0.0	$&$	0.0	$&$	 - 	$&$	 - 	$&$	0.0	$&$	0.0	$&$	 - 	$&$	 - 	$&$	0.0	$&$	0.0	$&$	 - 	$&$	 -  	$	\\
$\Im\CL^{XX}_{e\mu}$	&	$	 - 	$&$	 - 	$&$	-0.2	$&$	0.1	$&$	 - 	$&$	 - 	$&$	0.0	$&$	-0.1	$&$	 - 	$&$	 - 	$&$	0.0	$&$	-0.1	$&$	 - 	$&$	 - 	$&$	0.0	$&$	-0.1	$	\\
$\Im\CL^{XY}_{e\mu}$	&	$	 - 	$&$	 - 	$&$	0.2	$&$	-0.1	$&$	 - 	$&$	 - 	$&$	0.0	$&$	0.1	$&$	 - 	$&$	 - 	$&$	0.0	$&$	0.1	$&$	 - 	$&$	 - 	$&$	0.0	$&$	0.1	$	\\
$\Im\CL^{XZ}_{e\mu}$	&	$	0.4	$&$	-0.5	$&$	 - 	$&$	 - 	$&$	0.0	$&$	0.0	$&$	 - 	$&$	 - 	$&$	0.0	$&$	0.0	$&$	 - 	$&$	 - 	$&$	0.0	$&$	0.0	$&$	 - 	$&$	 -  	$	\\
$\Im\CL^{YY}_{e\mu}$	&	$	 - 	$&$	 - 	$&$	0.1	$&$	0.4	$&$	 - 	$&$	 - 	$&$	-0.2	$&$	0.0	$&$	 - 	$&$	 - 	$&$	-0.1	$&$	0.0	$&$	 - 	$&$	 - 	$&$	-0.2	$&$	0.0	$	\\
$\Im\CL^{YZ}_{e\mu}$	&	$	-0.5	$&$	-0.4	$&$	 - 	$&$	 - 	$&$	0.0	$&$	0.0	$&$	 - 	$&$	 - 	$&$	0.0	$&$	0.0	$&$	 - 	$&$	 - 	$&$	0.0	$&$	0.0	$&$	 - 	$&$	 -  	$	\\
\hline
Sensitivity &\multicolumn{4}{c|}{$<2\times10^{-23}$} & \multicolumn{4}{c|}{$<7\times10^{-23}$}&\multicolumn{4}{c|}{$<2\times10^{-23}$} & \multicolumn{4}{c}{$<2\times10^{-23}$}  \\
\end{tabular} 
\caption{\label{Table B}
Estimated amplitudes of sidereal-variation probabilities
$\PAsn{1}{e\mu}$, $\PAcn{1}{e\mu}$, 
$\PBsn{1}{e\mu}$, and $\PBcn{1}{e\mu}$
for appearance experiments
with $\nu_\mu\rightarrow\nu_e$.
The numerical value is listed for the estimated contribution 
to each amplitude from the real and imaginary parts
of the combinations
$\AL^\al_{e\mu}$ and $\CL^\ab_{e\mu}$
of fundamental coefficients for Lorentz and CPT violation.
Values are given to one decimal place,
in dimensionless units for $\AL^\al_{e\mu}$ 
and in units of GeV for $\CL^\ab_{e\mu}$.
A value of $0.0$ indicates rounding to zero at this precision,
while a dash implies the value is identically zero.
The last row lists the approximate sensitivity in GeV of each experiment.
} 
\end{center} 
\end{table*}

\subsubsection{Example: $\nu_e$ appearance}
\label{sec app}

Consider first searches for Lorentz and CPT violation
via $\nu_e$ appearance in a $\nu_\mu$ beam,
within the assumption of a comparatively large $\th_{13}$.
This requires a full three-flavor analysis.
We incorporate effects from matter-induced mixing 
via Eq.\ \rf{HSMmatter},
and we adopt explicit parameter values for vacuum mixing 
compatible with the observed three-flavor oscillations 
in solar, atmospheric, reactor, and accelerator experiments 
\cite{pdg},
\bea
\De m_\odot^2 & \simeq & 8.0\times10^{-5}\,\text{eV}^2 ,
\nonumber \\
\De m_\text{atm}^2& \simeq & 2.5\times10^{-3}\,\text{eV}^2 ,
\nonumber \\
\th_{12} \simeq  34^\circ ,
\quad
\th_{23} & \simeq & 45^\circ , 
\quad
\th_{13} \simeq 12^\circ ,
\quad
\de \simeq 0^\circ .
\qquad
\label{theta23=45}
\eea
Together with the matter effects,
the values \rf{theta23=45}
determine the linear combinations
of coefficients $\aL^\al_{ab}$ and $\cL^\ab_{ab}$ 
relevant for any given experiment.
Note that both magnitudes and signs are significant.
For example,
the above values hold for a normal mass hierarchy.
Our general expressions for oscillation probabilities 
are valid for any magnitudes and signs,
but the numerical results for the illustrative examples below
assume the specific choices \rf{theta23=45}.
For a comprehensive exploration 
of the space of coefficients for Lorentz and CPT violation,
distinct analyses of the same data
must be performed for each acceptable choice of parameter values
and must be reported as such. 
Note also that the approximation $\de\simeq 0^\circ$ 
made in Eq.\ \rf{theta23=45}
implies that there is little or no CP violation 
in standard oscillations,
although perturbative CP violation 
from Lorentz and CPT violation can still appear.

The anisotropies introduced by nonzero 
coefficients for Lorentz violation
can lead to sidereal variations,
which are the primary signals of interest here.
The sidereal decomposition of the probability 
takes the form \rf{P1sid}
with $\{ab\} = \{e\mu\}$.
The four amplitudes
$\PAsn{1}{e\mu}$,
$\PAcn{1}{e\mu}$,
$\PBsn{1}{e\mu}$,
and $\PBcn{1}{e\mu}$
are linear combinations of coefficients for Lorentz violation
given by Eqs.\ \rf{H1Bc} and \rf{P1Bc},
and they can be measured by studying
the variations of the neutrino mixing with sidereal time.
The effects from the combination
$\PCn{1}{e\mu}$ are more challenging to detect experimentally
because they have no accompanying time variation.

The amplitudes of the sidereal-variation probabilities
depend on the quantities
$\AL^\al_{e\mu}$ and $\CL^\ab_{e\mu}$,
which are linear combinations 
of fundamental coefficients for Lorentz and CPT violation,
\bea
\AL^\al_{e\mu} &=& \sum_{cd} \M{1}{e\mu}{cd}\, \aL^\al_{cd} ,
\nonumber\\
\CL^{\ab}_{e\mu} &=& \sum_{cd} \M{1}{e\mu}{cd}\, (c_L)^{\ab}_{cd} .
\label{c tilde}
\eea
The results of an experimental analysis
can therefore be expressed in
terms of the fundamental coefficients
$\aL^\al_{cd}$ and $\cL^\ab_{cd}$ 
by calculating the relevant complex factors $\M{1}{e\mu}{cd}$ 
for the given experiment.

Table \ref{Table A} presents approximate numerical values 
of the real and imaginary parts
of these factors
for the eight long-baseline experiments 
K2K, MINOS, OPERA, ICARUS, NO$\nu$A, T2K, DUSEL, and T2KK.
The entries are obtained assuming
the parameter values \rf{theta23=45}
and incorporating effects of matter-induced mixing.
The factors vary with the experiment,
reflecting their dependence on the baseline $L$
and the neutrino energy $E$.
Since the experiments 
are clustered in a single region of $L$-$E$ space,
the corresponding numerical values 
for each factor are roughly comparable 
and so the four sets of $\AL^\al_{e\mu}$, $\CL^\ab_{e\mu}$
obtained from Eq.\ \rf{c tilde}
are roughly comparable combinations
of the fundamental coefficients 
$\aL^\al_{ab}$, $\cL^\ab_{ab}$.
The table reveals that each experiment can measure 
particular combinations of
$\AL^\al_{e\mu}$ and $\CL^\ab_{e\mu}$,
which can then be used to constrain
the coefficient space of
$\aL^\al_{ab}$ and $\cL^\ab_{ab}$.
The large number of coefficients for Lorentz and CPT violation
and the limited number of observables for a given experiment
imply that multiple experiments are needed
to explore the entire coefficient space.
In performing the analysis,
it is of practical value to obtain an estimated maximal sensitivity
to each individual component of the fundamental coefficients
$\aL^\al_{ab}$ and $\cL^\ab_{ab}$ in turn,
by allowing only that component to be nonzero
and using the data to constrain it.

We can use Eqs.\ \rf{H1Bc} and \rf{P1Bc}
to calculate estimated first-order sensitivities to 
$\AL^\al_{e\mu}$ and $\CL^\ab_{e\mu}$
for each of these experiments.
Table \ref{Table B} lists the four amplitudes
$\PAsn{1}{e\mu}$,
$\PAcn{1}{e\mu}$,
$\PBsn{1}{e\mu}$,
and $\PBcn{1}{e\mu}$
as explicit linear combinations of the real and imaginary parts
of $\AL^\al_{e\mu}$ and $\CL^\ab_{e\mu}$.
The linear combinations typically differ for each experiment
and for each amplitude
because they depend on the beam energy $E$ 
and direction $\hat p$.
Note that the two experiments OPERA and ICARUS
have approximately the same baseline, orientation, and energy,
so they can be listed together for our purposes.
As an example,
the table reveals that in GeV units 
the amplitude $\PAsn{1}{e\mu}$ 
for the K2K experiment is approximately
\bea
\PAsn{1}{e\mu} &\approx&
-0.1 \Re\AL^{X}_{e\mu} +0.4\Re\CL^{TX}_{e\mu}
\nonumber\\
&&
+0.1\Im\AL^{X}_{e\mu} -0.3\Im\CL^{TX}_{e\mu} .
\eea
Using Table \ref{Table A},
all these combinations can be written in terms of 
the fundamental coefficients $\aL^\al_{ab}$ and $\cL^\ab_{ab}$ 
for Lorentz and CPT violation.

To obtain a crude estimate of the sensitivities 
for each experiment,
we suppose that a 10\% sidereal variation 
in the oscillation probability can be detected.
This leads to a sensitivity of order $10\%/2L$.
The last row of Table \ref{Table B}
lists these values for each experiment.
In conjunction with the other entries in Table \ref{Table B}
and with the factors listed in Table \ref{Table A},
these values can be used to obtain
the estimated first-order reach for any desired 
coefficient for Lorentz and CPT violation.

Table \ref{Table B} shows that long-baseline experiments
have the potential to achieve extreme sensitivities
to Lorentz and CPT violation
in the $\nu_\mu\to\nu_e$ appearance mode
if there is appreciable mass mixing
arising from a comparatively large value of $\th_{13}$.
Since some of the predicted effects are small,
second-order effects may also play a role
and may need to be incorporated
in a comprehensive analysis of real data.
If instead $\th_{13}$ is tiny or zero,
studies of Lorentz and CPT violation 
in the $\nu_\mu\to\nu_e$ appearance mode
can be performed using the methodology of Ref.\ \cite{km3},
as discussed in the previous subsection.

In the event that the listed experiments
run in antineutrino mode,
the attainable reach can be estimated similarly.
In the vacuum,
the factors $\M{1}{\bar a\bar b}{\bar c\bar d}$
are unaffected  
because the parameter values \rf{theta23=45}
imply CP invariance,
so the corresponding amplitudes 
of the sidereal-variation probabilities
can be found by replacing $\AL^\al_{e\mu}$ and $\CL^\ab_{e\mu}$
with $\AR^\al_{\bar e\bar \mu}$ and $\CR^\ab_{\bar e\bar \mu}$.
However,
the contribution from mass-induced mixing
in Eq.\ \rf{HSMmatter} enters with opposite sign,
so the estimated amplitudes in Table \ref{Table B}
acquire corresponding changes.

\subsubsection{Example: $\nu_\mu$ disappearance}
\label{sec 2nu}

\begin{table*}
\begin{center}
\begin{tabular}{c|cccc|cccc|cccc}
Experiment&\multicolumn{4}{c|}{K2K} & \multicolumn{4}{c|}{MINOS} &\multicolumn{4}{c}{OPERA, ICARUS} \\
\hline
Amplitude&
$\PASN{1}$ & $\PACN{1}$ & $\PBSN{1}$ & $\PBCN{1}$ &
$\PASN{1}$ & $\PACN{1}$ & $\PBSN{1}$ & $\PBCN{1}$ &
$\PASN{1}$ & $\PACN{1}$ & $\PBSN{1}$ & $\PBCN{1}$ \\ &&&&&&&&&&&\\[-10pt]
\hline
$\Re(a_L)^X_{\mu\tau}$	&$-0.5	$&$0.0	$&$ - 	$&$ - 	$&$-0.2	$&$0.2	$&$ - 	$&$ - 	$&$0.1	$&$-0.1	$&$ - 	$&$ - 	$\\
$\Re(a_L)^Y_{\mu\tau}$	&$0.0	$&$0.5	$&$ - 	$&$ - 	$&$0.2	$&$0.2	$&$ - 	$&$ - 	$&$-0.1	$&$-0.1	$&$ - 	$&$ - 	$\\
$\Re(c_L)^{TX}_{\mu\tau}$&$1.2	$&$-0.1	$&$ - 	$&$ - 	$&$0.7	$&$-1.0	$&$ - 	$&$ - 	$&$-3.5	$&$2.0	$&$ - 	$&$ - 	$\\
$\Re(c_L)^{TY}_{\mu\tau}$&$-0.1	$&$-1.2	$&$ - 	$&$ - 	$&$-1.0	$&$-0.7	$&$ - 	$&$ - 	$&$2.0	$&$3.5	$&$ - 	$&$ - 	$\\
$\Re(c_L)^{XX}_{\mu\tau}$&$- 	$&$ - 	$&$0.0	$&$0.3	$&$ - 	$&$ - 	$&$0.2	$&$-0.1	$&$ - 	$&$ - 	$&$0.8	$&$0.5	$\\
$\Re(c_L)^{XY}_{\mu\tau}$&$- 	$&$ - 	$&$0.0	$&$-0.3	$&$ - 	$&$ - 	$&$-0.2	$&$0.1	$&$ - 	$&$ - 	$&$-0.8	$&$-0.5	$\\
$\Re(c_L)^{XZ}_{\mu\tau}$&$-0.1	$&$0.0	$&$ - 	$&$ - 	$&$-0.4	$&$0.6	$&$ - 	$&$ - 	$&$-1.4	$&$0.8	$&$ - 	$&$ - 	$\\
$\Re(c_L)^{YY}_{\mu\tau}$&$- 	$&$ - 	$&$0.6	$&$-0.1	$&$ - 	$&$ - 	$&$-0.1	$&$-0.4	$&$ - 	$&$ - 	$&$0.9	$&$-1.6	$\\
$\Re(c_L)^{YZ}_{\mu\tau}$&$0.0	$&$0.1	$&$ - 	$&$ - 	$&$0.6	$&$0.4	$&$ - 	$&$ - 	$&$0.8	$&$1.4	$&$ - 	$&$ - 	$\\
\hline
Sensitivity &\multicolumn{4}{c|}{$<8\times10^{-23}$} & \multicolumn{4}{c|}{$<3\times10^{-23}$}&\multicolumn{4}{c}{$<3\times10^{-23}$}  \\
\end{tabular}
\vskip 20pt
\begin{tabular}{c|cccc|cccc|cccc|cccc}
Experiment&\multicolumn{4}{c|}{NO$\nu$A} & \multicolumn{4}{c|}{T2K} &\multicolumn{4}{c|}{DUSEL} & \multicolumn{4}{c}{T2KK} \\
\hline
Amplitude&
$\PASN{1}$ & $\PACN{1}$ & $\PBSN{1}$ & $\PBCN{1}$ &
$\PASN{1}$ & $\PACN{1}$ & $\PBSN{1}$ & $\PBCN{1}$ &
$\PASN{1}$ & $\PACN{1}$ & $\PBSN{1}$ & $\PBCN{1}$ &
$\PASN{1}$ & $\PACN{1}$ & $\PBSN{1}$ & $\PBCN{1}$ \\ &&&&&&&&&&&&&&&\\[-10pt]
\hline
$\Re(a_L)^X_{\mu\tau}$	&$-0.1	$&$0.2	$&$ - 	$&$ - 	$&$-0.2	$&$0.0	$&$ - 	$&$ - 	$&$-0.5	$&$0.0	$&$ - 	$&$ - 	$&$-0.5	$&$0.0	$&$ - 	$&$ -	$\\
$\Re(a_L)^Y_{\mu\tau}$	&$0.2	$&$0.1	$&$ - 	$&$ - 	$&$0.0	$&$0.2	$&$ - 	$&$ - 	$&$0.0	$&$0.5	$&$ - 	$&$ - 	$&$0.0	$&$0.5	$&$ - 	$&$ -	$\\
$\Re(c_L)^{TX}_{\mu\tau}$&$0.6	$&$-0.8	$&$ - 	$&$ - 	$&$0.2	$&$0.0	$&$ - 	$&$ - 	$&$1.0	$&$0.0	$&$ - 	$&$ - 	$&$0.8	$&$0.1	$&$ - 	$&$ -	$\\
$\Re(c_L)^{TY}_{\mu\tau}$&$-0.8	$&$-0.6	$&$ - 	$&$ - 	$&$0.0	$&$-0.2	$&$ - 	$&$ - 	$&$0.0	$&$-1.0	$&$ - 	$&$ - 	$&$0.1	$&$-0.8	$&$ - 	$&$ -	$\\
$\Re(c_L)^{XX}_{\mu\tau}$&$ - 	$&$ - 	$&$0.2	$&$-0.1	$&$ - 	$&$ - 	$&$0.0	$&$0.1	$&$ - 	$&$ - 	$&$0.0	$&$0.2	$&$ - 	$&$ - 	$&$0.0	$&$0.2	$\\
$\Re(c_L)^{XY}_{\mu\tau}$&$ - 	$&$ - 	$&$-0.2	$&$0.1	$&$ - 	$&$ - 	$&$0.0	$&$-0.1	$&$ - 	$&$ - 	$&$0.0	$&$-0.2	$&$ - 	$&$ - 	$&$0.0	$&$-0.2	$\\
$\Re(c_L)^{XZ}_{\mu\tau}$&$-0.4	$&$0.5	$&$ - 	$&$ - 	$&$0.0	$&$0.0	$&$ - 	$&$ - 	$&$-0.2	$&$0.0	$&$ - 	$&$ - 	$&$0.1	$&$0.0	$&$ - 	$&$ -	$\\
$\Re(c_L)^{YY}_{\mu\tau}$&$ - 	$&$ - 	$&$-0.1	$&$-0.3	$&$ - 	$&$ - 	$&$0.1	$&$0.0	$&$ - 	$&$ - 	$&$0.5	$&$0.0	$&$ - 	$&$ - 	$&$0.4	$&$0.1	$\\
$\Re(c_L)^{YZ}_{\mu\tau}$&$0.5	$&$0.4	$&$ - 	$&$ - 	$&$0.0	$&$0.0	$&$ - 	$&$ - 	$&$0.0	$&$0.2	$&$ - 	$&$ - 	$&$0.0	$&$-0.1	$&$ - 	$&$ -	$\\
\hline
Sensitivity &\multicolumn{4}{c|}{$<2\times10^{-23}$} & \multicolumn{4}{c|}{$<7\times10^{-23}$}&\multicolumn{4}{c|}{$<2\times10^{-23}$} & \multicolumn{4}{c}{$<2\times10^{-23}$}  \\
\end{tabular} 
\caption{\label{Table D}
Estimated amplitudes of sidereal-variation probabilities
$\PAsn{1}{\mu\ta}$, $\PAcn{1}{\mu\ta}$, 
$\PBsn{1}{\mu\ta}$, and $\PBcn{1}{\mu\ta}$
within the two-generation approximation.
The numerical value is listed for the estimated contribution 
to each amplitude from the real parts
of the fundamental coefficients 
$\aL^\al_{\mu\ta}$ and $\cL^\ab_{\mu\ta}$
for Lorentz and CPT violation.
Values are given to one decimal place,
in dimensionless units for $\aL^\al_{\mu\ta}$ 
and in units of GeV for $\cL^\ab_{\mu\ta}$.
A value of $0.0$ indicates rounding to zero at this precision,
while a dash implies the value is identically zero.
The last row lists the approximate sensitivity in GeV of each experiment.
Analogous results for possible experiments with antineutrinos 
can be obtained by replacing 
$\aL^\al_{\mu\ta}$ and $\cL^\ab_{\mu\ta}$
with $\aR^\al_{\bar \mu\bar \ta}$ and $\cR^\ab_{\bar \mu\bar \ta}$.} 
\end{center} 
\end{table*}

Beams of $\nu_\mu$ also provide opportunities
to search for $\nu_\mu$ disappearance.
The probability of $\nu_\mu$ oscillating into
other neutrinos is given by
\beq
P_{\nu_\mu\rightarrow\nu_X}=1-P_{\nu_\mu\rightarrow\nu_\mu}.
\eeq
The correction introduced
by $\aL^\al_{ab}$ and $\cL^\ab_{ab}$ coefficients
is therefore given by
\beq 
P_{\nu_\mu\rightarrow\nu_X}^{(1)}
= -P_{\nu_\mu\rightarrow\nu_\mu}^{(1)} 
= -2L\,\Im\big((S_{\mu\mu}^{(0)})^*\,\H{1}{\mu\mu}\big) ,
\eeq
where $\H{1}{\mu\mu}$ is defined in Eq.\ \rf{formone}.
Again,
sidereal variations can arise 
from the anisotropies introduced by Lorentz violation,
and $\H{1}{\mu\mu}$ can be expanded in sidereal time
according to Eq.\ \rf{H1sid}.

For $\nu_\mu$ disappearance experiments,
the sidereal decomposition of the probability 
is given by Eq.\ \rf{P1sid}
with $\{ab\} = \{\mu\mu\}$.
The four amplitudes 
$\PAsn{1}{\mu\mu}$,
$\PAcn{1}{\mu\mu}$,
$\PBsn{1}{\mu\mu}$,
$\PBcn{1}{\mu\mu}$
of the sidereal-variation probabilities
are provided in Eqs.\ \rf{H1Bc} and \rf{P1Bc}.
Their exact expressions in terms of the combinations
$\AL^\al_{\mu\mu}$ and $\CL^\ab_{\mu\mu}$
of fundamental coefficients for Lorentz and CPT violation
take the form
\bea
\AL^\al_{\mu\mu} &=& 
\sum_{cd} \M{1}{\mu\mu}{cd}\, \aL^\al_{cd} ,
\nonumber\\
\CL^{\ab}_{\mu\mu} &=& 
\sum_{cd} \M{1}{\mu\mu}{cd}\, (c_L)^{\ab}_{cd} .
\label{mumu}
\eea
If needed,
the experiment-dependent complex factors $\M{1}{\mu\mu}{cd}$
controlling these linear combinations
can be obtained using Eq.\ \rf{mone}.

In scenarios with oscillations occurring primarily 
between $\nu_\mu$ and $\nu_\ta$,
the probability for $\nu_\mu$ disappearance
can be well approximated 
by restricting attention to two-flavor vacuum mixing.
This limit involves only one mass-squared difference 
and one mixing angle,
and it offers another useful illustration
of the general analysis given in Sec.\ \ref{sec osc}.
In the event of tiny or zero $\th_{13}$,
it is also the relevant limit
for the eight experiments considered above.
We adopt this case as our second illustrative example.

In the Lorentz-invariant two-flavor limit,
an overall diagonal term can be removed
from the hamiltonian $(h_0)_{ab}$
because it is irrelevant for oscillations.
This gives a 2$\times$2 mass matrix of the form
\beq
(h_0)_{ab}
\simeq \frac{1}{2E}U^\dag \begin{pmatrix} 0 & 0 \\
0 & \Delta m^2_{32} \end{pmatrix}U .
\label{H_2nu}  
\eeq
The flavor indices are now restricted to two generations,
$a,b,\ldots = \mu,\ta$.
The mixing matrix $U$ depends on the mixing angle $\th_{23}$
according to
\beq 
U_{a'a}=\begin{pmatrix} c_{23} & -s_{23} \\
s_{23} & c_{23} \end{pmatrix} .
\eeq
Note that CP violation due to mass mixing 
is strictly unobservable in this limit.
The single mass difference is
then given by
\beq
\De m^2_{32} = \De m^2_{\text{atm}} - \De m^2_{\odot}
\simeq \De m^2_{\text{atm}} .
\eeq
For the explicit estimations in this subsection,
we choose for 
$\De m^2_{\text{atm}}$ and $\th_{23}$
the values
\bea
\De m_\text{atm}^2& \simeq & 2.5\times10^{-3}\,\text{eV}^2 ,
\nonumber \\
\th_{23} & \simeq & 45^\circ ,
\label{twoflavorparm}
\eea
which are consistent with the three-flavor 
parameter values \rf{theta23=45}.

In the presence of Lorentz and CPT violation,
the two-generation approximation simplifies
the expression \rf{P1na}
for the first-order oscillation probabilities.
With the two flavors being $\nu_\mu$ and $\nu_\ta$,
we have 
\beq
\S{0}{e\mu}=\S{0}{e\ta}=\S{0}{\mu e}=\S{0}{\ta e}=0.
\eeq
This implies no mixing with electron neutrinos 
occurs in the first-order perturbation.
Also, 
inspection of the form of
$\M{1}{ab}{cd}(t)$ given in Eq.\ \rf{mone}
reveals that only those coefficients for Lorentz violation
lying in the two-flavor $\{\mu\ta\}$ subspace 
can lead to first-order effects.

Explicitly,
we find the first-order oscillation probabilities are 
\bea
P_{\nu_\mu\mix\nu_\tau}^{(1)}
&=& -P_{\nu_\mu\to\nu_\mu}^{(1)}
= -P_{\nu_\tau\to\nu_\tau}^{(1)}
\nonumber \\
&=& 2L\,\Im\big((S_{\tau\mu}^{(0)})^*\,\H{1}{\tau\mu}\big)
\nonumber\\
&\approx & \Re(\de h_{\mu\tau})L\,\sin{(\Delta m^2_{32}L/2E)} .
\label{2nu prob}
\eea
We assume maximal mixing in the last expression,
in accordance with the parameter values \rf{twoflavorparm}.
Note that only the real part of $(\de h)_{\mu\ta}$
contributes to first-order mixing in this limit.
Also, 
the corresponding antineutrino mixing probabilities
are found by the index replacements
$\{\mu\ta\}\to\{\bar\mu\bar\ta\}$,
which is equivalent to changing
the sign of the coefficient $\aL^\al_{\mu\ta}$ in $(\de h)_{\mu\ta}$.

The oscillation probability \rf{2nu prob}
can be decomposed into sidereal amplitudes
according to Eq.\ \rf{P1sid}.
The amplitudes take the form \rf{P1Bc}
with the definitions \rf{H1Bc}.
The two-flavor approximation makes it straightforward
to express the latter directly
in terms of the real parts of
the fundamental coefficients
$\aL^\al_{\mu\ta}$ and $\cL^\ab_{\mu\ta}$
rather than the intermediate combinations 
$\AL^\al_{\mu\ta}$ and $\CL^\ab_{\mu\ta}$.
We can use these results 
to estimate experimental sensitivities
to the real parts of $\aL^\al_{\mu\ta}$ and $\cL^\ab_{\mu\ta}$
for any specified experiment.

Table \ref{Table D} presents the results 
of estimates for the eight long-baseline beam experiments
considered above.
In each experiment, 
numerical values are listed for the weighting
of the real parts $\aL^\al_{\mu\ta}$ and $\cL^\ab_{\mu\ta}$
in the four amplitudes
$\PAsn{1}{\mu\ta}$,
$\PAcn{1}{\mu\ta}$,
$\PBsn{1}{\mu\ta}$,
and $\PBcn{1}{\mu\ta}$.
The last row provides a rough approximation
to the attainable sensitivity,
based on assuming that a 10\% sidereal variation
in the oscillation probability can be detected.
The results indicate that all these experiments
can achieve impressive sensitivities 
to perturbative Lorentz and CPT violation.

\subsection{CP and CPT asymmetries}
\label{sec cpt}

In this subsection,
we discuss the effects of Lorentz violation
on tests of the discrete symmetries CP and CPT
using neutrino oscillations.
Experimentally,
nature is known to break CP invariance
in the weak interactions,
although no CP violation in neutrino oscillations
has yet been detected.
In contrast,
compelling evidence for violation of CPT symmetry 
in any system is lacking to date.
On the theoretical front,
CPT invariance has a profound connection
to Lorentz invariance in quantum field theory,
where the CPT theorem shows that 
under mild assumptions
CPT violation is accompanied by Lorentz violation
\cite{owg}.
No such relationship exists for CP, 
and indeed CP may be violated even when Lorentz invariance holds.

Consider first CP violation.
The CP transformation interchanges 
oscillation probabilities according to 
\beq
P_{\nu_a\to\nu_b}\stackrel{\rm CP}{\mix} P_{\nub_a\to\nub_b} ,
\eeq
so CP violation can be revealed as differences 
in neutrino and antineutrino probabilities.
A generic measure of CP violation 
for mixing involving flavors $\{a,b\}$
is the asymmetry 
\beq
\cA^{CP}_{ab} = \frac
{P_{\nu_a\to\nu_b}- P_{\nub_a\to\nub_b}}
{P_{\nu_a\to\nu_b}+ P_{\nub_a\to\nub_b}} ,
\label{cpasym}
\eeq
which is zero when CP is a symmetry of the oscillations.

Similar results hold for CPT.
Under the CPT transformation,
the probabilities exchange according to
\beq
P_{\nu_a\to\nu_b}\stackrel{\rm CPT}{\mix} P_{\nub_b\to\nub_a} .
\eeq
For mixing involving flavors $\{a,b\}$,
we can therefore define the asymmetry 
\beq
\cA^{CPT}_{ab} = \frac
{P_{\nu_a\to\nu_b}- P_{\nub_b\to\nub_a}}
{P_{\nu_a\to\nu_b}+ P_{\nub_b\to\nub_a}} .
\label{cptasym}
\eeq
This asymmetry vanishes if CPT invariance holds.
Note,
however,
that the converse is false:
models can be constructed in which CPT is violated
even when the asymmetry $\cA^{CPT}_{ab}$ vanishes
\cite{km1}.
In such cases,
detailed studies of energy and direction dependences 
may be required to reveal CPT violation.

For the special case of two-flavor models,
the probabilities are blind to possible $T$ violation,
so $P_{\nu_a \to \nu_b} = P_{\nu_b \to \nu_a}$
and 
$P_{\nub_a \to \nub_b} = P_{\nub_b \to \nub_a}$.
Consequently,
the CP and CPT asymmetries 
are identical in the two-flavor limit,
\beq
\cA^{CP}_{ab} = \cA^{CPT}_{ab}, 
\quad a,b = \mu,\ta.
\eeq
Note,
however,
that CPT violation may present itself in other ways,
such as unconventional energy and direction dependence.

The above asymmetries can be used to test both CP and CPT 
in neutrino-oscillation experiments.
In the Lorentz-invariant case,
violation of CP symmetry occurs 
when both the mixing angle $\th_{13}$ 
and the CP phase $\de$ are nonzero.
Measuring $\th_{13}$ and searching for CP violation
are major goals of many forthcoming oscillation experiments.
Some experiments can change polarity,
choosing to focus either positively or negatively charged mesons 
into the decay pipe,
and hence can run in both neutrino and antineutrino modes. 
This feature may permit high-statistics direct searches 
for CP violation.
The nature of the beam or other properties may also
lead to accumulation of neutrino and antineutrino data.
In all these cases,
both Lorentz-conserving and Lorentz-violating situations
can be accessed,
thereby enabling also searches for CPT violation.

The interpretation of asymmetries constructed 
from experimental data requires a theoretical framework.
One phenomenological approach to CPT violation 
assumes different masses and mixing angles 
for neutrinos and antineutrinos.
In the two-flavor case,
for example,
this approach takes a set of parameters 
$(\De m^2,\th)$ for neutrinos
and a second set
$(\De \ol m^2,\ol \th)$
for antineutrinos.
It is tempting to adopt the resulting explicit expression 
for the asymmetry $\cA^{CPT}_{ab}\equiv \cA^{CP}_{ab}$
for purposes of data analysis and interpretation,
treating the parameters
$\De m^2$, $\De \ol m^2$, $\th$, and $\ol \th$
as Lorentz-scalar constants.
However,
according to the CPT theorem
this procedure is inconsistent with quantum field theory
because under mild assumptions CPT violation in field theory
must come with Lorentz violation 
\cite{owg},
so the parameters $\De m^2$, $\De \ol m^2$, $\th$, and $\ol\th$
cannot be Lorentz scalars.
Instead,
they must depend on the 4-momentum of the neutrino,
including both the energy $E$ and the propagation direction $\hat p$
relative to the Sun-centered frame.
A typical experiment involves neutrinos spanning
a spectrum of values for $E$ and $\hat p$.
The 4-momentum dependence of the asymmetry
therefore entails significant consequences
for data analysis and its interpretation 
in searches for CPT violation.

As an illustration,
we derive here the explicit first-order form 
of the two-flavor CPT asymmetries
$\cA^{CPT}_{\mu\ta} = \cA^{CP}_{\mu\ta}$ 
and 
$\cA^{CPT}_{\mu\mu} = \cA^{CP}_{\mu\mu}$ 
in the field-theoretic context.
For definiteness we assume maximal mixing,
which is consistent with the parameter values \rf{twoflavorparm}.
At first order,
calculation reveals these asymmetries
depend on the coefficients $\aL^\al_{\mu\ta}$
for Lorentz and CPT violation
but are independent of $\cL^\ab_{\mu\ta}$.

To present the asymmetries,
it is convenient to
introduce the CPT-odd part $(\de h)_{\mu\ta}^{CPT}$
of the perturbative hamiltonian $(\de h)_{\mu\ta}$ 
with coefficients expressed in the Sun-centered frame,
\bea
(\de h)_{\mu\ta}^{CPT}
&\equiv&
(\de h)_{\mu\ta}\big\vert_{c_L\to 0}
\nonumber\\
&=&
\aL^T_{\mu\ta}-\Nh^Z \aL^Z_{\mu\ta} 
\nonumber \\
&&
+\big( \Nh^Y \aL^X_{\mu\ta}-\Nh^X\, 
\aL^Y_{\mu\ta} \big) \sin \om_\oplus T_\oplus 
\nonumber\\
&&
-\big(\Nh^X\aL^X_{\mu\ta}+\Nh^Y\, 
\aL^Y_{\mu\ta} \big) \cos \om_\oplus T_\oplus .
\nonumber\\
\label{delham}
\eea
In terms of this quantity,
we find that the CPT asymmetry $\cA^{CPT}_{\mu\ta}$ is 
\bea
\cA^{CPT}_{\mu\ta} &=& \cA^{CP}_{\mu\ta}\approx  
2L\, \cot\big( \frac{\De m^2_{32}L}{4E}\big) 
\Re (\de h)_{\mu\ta}^{CPT} .
\qquad
\label{2nu ratio}
\eea
This result is valid provided the experiment 
operates away from the region of parameter space
leading to small oscillations,
$\sin\big(\De m^2_{32}L/4E\big)\ /\hspace{-10pt}\approx 0$.
For the second CPT asymmetry
$\cA^{CPT}_{\mu\mu}$,
we obtain 
\bea
\cA^{CPT}_{\mu\mu} &=& \cA^{CP}_{\mu\mu}\approx 
-2L\, \tan\big(\frac{\De m^2_{32}L}{4E}\big) 
\Re (\de h)_{\mu\ta}^{CPT} ,
\qquad
\label{mumuratio}
\eea
where now we assume the experiment 
operates away from the region of parameter space
leading to large oscillations,
$\sin\big(\De m^2_{32}L/4E\big)\ /\hspace{-10pt}\approx 1$.
Inspection of these results reveals that
the two asymmetries 
$\cA^{CPT}_{\mu\ta}$ in Eq.\ \rf{2nu ratio}
and $\cA^{CPT}_{\mu\mu}$ in Eq.\ \rf{mumuratio}
contain the same essential information
about CPT violation
but are valid in different regions of parameter space.
In practice,
at least one of the two asymmetries
can be applied for a given experiment.

The results \rf{2nu ratio} and \rf{mumuratio}
display several interesting features.
The asymmetries grow with baseline $L$,
so experiments with comparable statistical power
but longer baselines have improved sensitivity. 
According to Eq.\ \rf{delham},
the asymmetries also vary with sidereal time $T_\oplus$
and depend on the direction of the neutrino beam.
Both these effects are features of CPT violation
and its accompanying Lorentz breaking.
We remark in passing that the structure
of the above equations bears a close similarity
to that of the analogous measures for CPT violation
in studies of neutral mesons.
For example,
the dependence of $(\de h)_{\mu\ta}^{CPT}$
on the coefficients $\aL^\al_{\mu\ta}$
in Eq.\ \rf{delham}
parallels that of the measure of CPT violation
given in Eq.\ (14) of Ref.\ \cite{ak3}. 

In a given experiment,
measuring the amplitudes of the sidereal variations
in the asymmetries \rf{2nu ratio} and \rf{mumuratio}
may produce interesting sensitivities
to the coefficient combinations
$\big( \Nh^Y \aL^X_{\mu\ta}-\Nh^X \aL^Y_{\mu\ta} \big)$
and $\big(\Nh^X\aL^X_{\mu\ta}+\Nh^Y \aL^Y_{\mu\ta} \big)$.
These combinations are independent 
of the coefficients $\cL^\ab_{\mu\ta}$
for CPT-even Lorentz violation.
Inspection of Eq.\ \rf{delham} reveals that
each asymmetry also depends on 
the coefficients $\aL^T_{\mu\ta}$ and $\aL^Z_{\mu\ta}$,
which are inaccessible via direct sidereal decomposition
of the oscillation probabilities or asymmetries.
One way to extract sensitivity to these coefficients
is to average the data over time,
in analogy to the extraction 
of the corresponding coefficients for CPT violation
in experiments with neutral mesons
\cite{ak12}.
The time-averaged asymmetry $\ol{\cA^{CPT}_{\mu\ta}}$ is
\bea
\ol{\cA^{CPT}_{\mu\ta}} &\approx&
2L\, \cot\big( \frac{\De m^2_{32}L}{4E}\big)  
\Re \big[ 
\aL^T_{\mu\ta}-\Nh^Z \aL^Z_{\mu\ta} 
\big] ,
\nonumber\\
\eea
while the time-averaged asymmetry $\ol{\cA^{CPT}_{\mu\mu}}$ is
\bea
\ol{\cA^{CPT}_{\mu\mu}} &\approx&
-2L\, \tan\big( \frac{\De m^2_{32}L}{4E}\big)  
\Re \big[ 
\aL^T_{\mu\ta}-\Nh^Z \aL^Z_{\mu\ta} 
\big] .
\nonumber\\
\eea
Note that these results remain dependent
on the beam direction despite the time averaging.
Each asymmetry therefore typically has distinct physical meanings 
for different experiments.
For example,
the directional factor $\Nh^Z$ is
$\Nh^Z \simeq 0.1$ for K2K,
$\Nh^Z \simeq 0.6$ for MINOS, 
$\Nh^Z \simeq -0.4$ for OPERA and ICARUS,
$\Nh^Z \simeq 0.6$ for NO$\nu$A,
$\Nh^Z \simeq -0.01$ for T2K,
$\Nh^Z \simeq 0.2$ for DUSEL,
and
$\Nh^Z \simeq -0.1$ for T2KK.  

While experiments capable of CP tests
necessarily test for CPT signals
in the two-flavor approximation,
the CPT signature 
$P_{\nu_a\to\nu_b}\neq P_{\nub_b\to\nub_a}$
may be more challenging to detect in three-neutrino scenarios.
Data from the accelerator experiments discussed above
or from next-generation studies using a beta beam
\cite{beta}
or a dedicated neutrino factory
\cite{nufact}
could be well suited for seeking three-flavor CPT violation
through direct comparisons of neutrinos and antineutrinos
using the asymmetry ${\cA^{CPT}_{ab}}$ of Eq.\ \rf{cptasym}.
Suitable comparisons of neutrinos and antineutrinos,
perhaps including time averaging as above,
could also lead to measurements of coefficient combinations 
without accompanying sidereal variations.

\section{Coefficients $\gt^\ab_{a\bar b}$ and $\Ht^\al_{a\bar b}$}
\label{sec gH}

In this section,
we discuss the dominant effects 
on neutrino oscillations arising from the coefficients
$\gt^\ab_{a\bar b}$ and $\Ht^\al_{a\bar b}$.
As shown in Eq.\ \rf{P1na},
these coefficients leave the oscillation probabilities
unaffected at first order.
The dominant effects appear at second order,
where the probabilities are given by Eq.\ \rf{P2an}.

The features introduced by 
$\gt^\ab_{a\bar b}$ and $\Ht^\al_{a\bar b}$
include unconventional energy and directional dependences.
However,
some key differences arise compared to the case 
of the coefficients $\aL^\al_{ab}$ and $\cL^\ab_{ab}$.
For example,
the dominant sidereal variations include higher harmonics
with frequencies up to $4\om_\oplus$.
Another example is mixing between neutrinos and antineutrinos
\cite{km1},
which violates lepton-number conservation.
This feature arises because 
$\gt^\ab_{a\bar b}$ and $\Ht^\al_{a\bar b}$
lie in the off-diagonal blocks 
of the perturbative hamiltonian \rf{dh}.

In Sec.\ \ref{sec lnv},
we focus on the second-order contributions
to the neutrino-antineutrino oscillation probability 
$P^{(2)}_{\nub_b\to\nu_a}$,
which involves lepton-number violation.
The second-order effects 
on the neutrino-neutrino and antineutrino-antineutrino 
mixing probabilities
$P^{(2)}_{\nu_b\to\nu_a}$ and 
$P^{(2)}_{\bar \nu_b\to\bar \nu_a}$
are considered in Sec.\ \ref{sec lnc}.

\subsection{Oscillations violating lepton number}
\label{sec lnv}

In this subsection,
we derive the sidereal behavior 
of the second-order oscillation probability
$P^{(2)}_{\nub_b\to\nu_a}$
for neutrino-antineutrino mixing.
Equations \rf{H1na} and \rf{P2an} 
specify this probability 
in terms of the perturbative hamiltonian
$\de h_{a\bar b}$,
which itself depends on the coefficients
$\gt^\ab_{a\bar b}$ and $\Ht^\al_{a\bar b}$
according to Eq.\ \rf{han}.
Note that neutrino-antineutrino oscillations 
are independent of the coefficients
$\aL^\al_{ab}$ and $\cL^\ab_{ab}$
at this order.

To determine the sidereal decomposition 
of the off-diagonal block $\de h_{a\bar b}$ 
of the perturbative hamiltonian,
we note that 
$\Ht^\al_{a\bar b}$ is an observer vector
and hence induces effects at frequency $\om_\oplus$,
while $\gt^\ab_{a\bar b}$ is a 2-tensor 
and hence induces effects at $\om_\oplus$ and $2\om_\oplus$.
The sidereal decomposition of $\de h_{a\bar b}$ 
therefore takes the form
\bea
\de h_{a\bar b} &\equiv&
-i\sqrt2 (\ep_+)_\al \big[\gt^\ab p_\be -\Ht^\al \big]_{a\bar b}
\nonumber \\
&=&\C{a\bar b}
+\As{a\bar b} \sin\om_\oplus T_\oplus
+\Ac{a\bar b} \cos\om_\oplus T_\oplus
\nonumber \\
&&+\Bs{a\bar b} \sin2\om_\oplus T_\oplus
+\Bc{a\bar b} \cos2\om_\oplus T_\oplus .
\label{habbarsid}
\eea
In this expression,
the amplitudes
$\C{a\bar b}$,
$\As{a\bar b}$,
$\Ac{a\bar b}$,
$\Bs{a\bar b}$, and
$\Bc{a\bar b}$
are direction-dependent linear combinations
of the coefficients 
$\gt^\ab_{a\bar b}$ and $\Ht^\al_{a\bar b}$
for Lorentz violation.

The direction dependence is governed by two vectors,
the momentum $\vec p$ and
the polarization $\vec\ep_+$.
The momentum is determined by the beam direction,
which varies sidereally and is specified 
at time $T_\oplus = 0$ in the Sun-centered frame 
by the vector $(\Nh^X,\Nh^Y,\Nh^Z)$
given in local spherical coordinates by Eq.\ \rf{nvector}.
We denote the analogous vector for $\vec\ep_+$
by $(\Ep^X,\Ep^Y,\Ep^Z)$.
In the same local spherical coordinates,
this vector has components
\bea
\Ep^X &=& \fr1{\sqrt2}\big(\cos\ch(\cos\th\cos\ph-i\sin\ph)
-\sin\ch\sin\th\def\emf{\EuScript{E}}\big) ,
\nonumber\\
\Ep^Y &=& \fr1{\sqrt2}\big(\cos\th\sin\ph+i\cos\ph\big) ,
\nonumber\\
\Ep^Z &=&-\fr1{\sqrt2}\big(\sin\ch(\cos\th\cos\ph-i\sin\ph)
+\cos\ch\sin\th\big) ,
\nonumber\\
\label{epvector}
\eea
where we have used the expression \rf{epplus}
for $(\ep_+)^j$.

Some calculation reveals that
the sidereal amplitudes in Eq.\ \rf{habbarsid}
are given as
\bea
\C{a\bar b} &=& -i\sqrt2\big[
\Ep^Z \Ht^Z_{a\bar b}
-\Ep^Z E \gt^{ZT}_{a\bar b}
\nonumber \\ 
&&
+\Ep^Z \Nh^Z E (\gt^{ZZ}_{a\bar b} - \half\gt^{XX}_{a\bar b} 
- \half\gt^{YY}_{a\bar b})
\nonumber \\ 
&&
+\fr{i}{2} \Ep^Z E (\gt^{XY}_{a\bar b} - \gt^{YX}_{a\bar b})
\big] ,
\nonumber \\ 
\As{a\bar b} &=& -i\sqrt2\big[
-\Ep^Y \Ht^X_{a\bar b}
+\Ep^X \Ht^Y_{a\bar b}
\nonumber \\ 
&&
+\Ep^Y E \gt^{XT}_{a\bar b}
-\Ep^X E \gt^{YT}_{a\bar b}
\nonumber \\ 
&&
-\Ep^Y \Nh^Z E \gt^{XZ}_{a\bar b}
+\Ep^X \Nh^Z E \gt^{YZ}_{a\bar b}
\nonumber \\ 
&&
-\Ep^Z \Nh^Y E \gt^{ZX}_{a\bar b}
+\Ep^Z \Nh^X E \gt^{ZY}_{a\bar b}
\big] ,
\nonumber \\ 
\Ac{a\bar b} &=& -i\sqrt2\big[
\Ep^X \Ht^X_{a\bar b}
+\Ep^Y \Ht^Y_{a\bar b}
\nonumber \\ 
&&
-\Ep^X E \gt^{XT}_{a\bar b}
-\Ep^Y E \gt^{YT}_{a\bar b}
\nonumber \\ 
&&
+\Ep^X \Nh^Z E \gt^{XZ}_{a\bar b}
+\Ep^Y \Nh^Z E \gt^{YZ}_{a\bar b}
\nonumber \\ 
&&
+\Ep^Z \Nh^X E \gt^{ZX}_{a\bar b}
+\Ep^Z \Nh^Y E \gt^{ZY}_{a\bar b}
\big] ,
\nonumber \\ 
\Bs{a\bar b} &=& -i\sqrt2\big[
\half(\Ep^X\Nh^X-\Ep^Y\Nh^Y) E (\gt^{XY}_{a\bar b}
+\gt^{YX}_{a\bar b})
\nonumber \\ 
&&
-\half(\Ep^X\Nh^Y+\Ep^Y\Nh^X) E (\gt^{XX}_{a\bar b}
-\gt^{YY}_{a\bar b})
\big] ,
\nonumber \\ 
\Bc{a\bar b} &=& -i\sqrt2\big[
\half(\Ep^X\Nh^Y+\Ep^Y\Nh^X) E (\gt^{XY}_{a\bar b}
+\gt^{YX}_{a\bar b})
\nonumber \\ 
&&
+\half(\Ep^X\Nh^X-\Ep^Y\Nh^Y) E (\gt^{XX}_{a\bar b}
-\gt^{YY}_{a\bar b})
\big] .
\label{h2Bc}
\eea
This completes the decomposition of $\de h_{a\bar b}$ 
in terms of the sidereal time $T_\oplus$,
the coefficients $\gt^\ab_{a\bar b}$, $\Ht^\al_{a\bar b}$,
and the components $\Nh^J$, $\Ep^J$.
The analogous decomposition for $\de h_{\bar ba}$ 
is obtained by taking the hermitian conjugate,
following Eq.\ \rf{han}.

At dominant order,
the neutrino-antineutrino mixing 
is controlled by the linear combinations
$\H{1}{a\bar b}$ and $\H{1}{\bar ab}$
given in Eq.\ \rf{H1na}.
The sidereal dependence of $\de h_{a\bar b}$ 
transfers to these combinations,
leading to the expansion
\bea
\H{1}{a\bar b} &=&
\Cn{1}{a\bar b}
\nonumber \\
&&
+\Asn{1}{a\bar b} \sin\om_\oplus T_\oplus
+\Acn{1}{a\bar b} \cos\om_\oplus T_\oplus
\nonumber \\
&&
+\Bsn{1}{a\bar b} \sin2\om_\oplus T_\oplus
+\Bcn{1}{a\bar b} \cos2\om_\oplus T_\oplus ,
\nonumber\\ 
\label{H1na sid}
\eea
with a similar expression for $\H{1}{\bar ab}$.
The coefficients
$\gt^\ab_{a\bar b}$ and $\Ht^\al_{a\bar b}$
appear in this expansion
in linear combinations weighted 
by the complex experiment-dependent factors
$\M{1}{a\bar b}{c\bar d}$ and
$\M{1}{\bar ab}{\bar cd}$
given in Eq.\ \rf{mone}.
It is convenient to introduce the definitions 
\bea
\gT^\ab_{a\bar b}&=&
\sum_{c\bar d} \M{1}{a\bar b}{c\bar d}\gt^\ab_{c\bar d} , 
\nonumber\\
\HT^\al_{a\bar b}&=&
\sum_{c\bar d} \M{1}{a\bar b}{c\bar d}\Ht^\al_{c\bar d} , 
\nonumber\\
\gT^\ab_{\bar ab}&=&
\sum_{\bar cd} \M{1}{\bar ab}{\bar cd}\gt^\ab_{\bar cd}
=\sum_{\bar cd} \M{1}{\bar ab}{\bar cd}\gt^{\ab\, *}_{d\bar c} , 
\nonumber\\
\HT^\al_{\bar ab}&=&
\sum_{\bar cd} \M{1}{\bar ab}{\bar cd}\Ht^\al_{\bar cd}
= \sum_{\bar cd} \M{1}{\bar ab}{\bar cd}\Ht^{\al\, *}_{d\bar c} .
\eea
In terms of these,
the sidereal amplitudes
$\Cn{1}{a\bar b}$,
$\Asn{1}{a\bar b}$,
$\Acn{1}{a\bar b}$,
$\Bsn{1}{a\bar b}$, 
$\Bcn{1}{a\bar b}$
take the same form as the corresponding amplitudes
in Eq.\ \rf{h2Bc}
but with 
$\gt^\ab_{a\bar b}$ and $\Ht^\al_{a\bar b}$
replaced with
$\gT^\ab_{a\bar b}$ and $\HT^\al_{a\bar b}$.
For the coefficient combinations $\H{1}{\bar ba}$,
we can define analogous sidereal amplitudes
$\Cn{1}{\bar ba}$,
$\Asn{1}{\bar ba}$,
$\Acn{1}{\bar ba}$,
$\Bsn{1}{\bar ba}$, 
$\Bcn{1}{\bar ba}$.
The forms of these can also be obtained from Eq.\ \rf{h2Bc},
by first taking the hermitian conjugates 
of the expressions on the right-hand side
and then replacing
$\gt^{\ab}_{\bar ba}$, $\Ht^{\ab}_{\bar ba}$
with
$\gT^\ab_{\bar ba}$, $\HT^\al_{\bar ba}$.

According to Eq.\ \rf{P2an},
the combinations $\H{1}{a\bar b}$
contribute quadratically 
to the second-order neutrino-antineutrino probabilities 
$P^{(2)}_{\nub_b\to\nu_a}$.
This implies that sidereal variations 
at frequencies up to $4\om_\oplus$ are observable.
Consequently, 
we expand $P^{(2)}_{\nub_b\to\nu_a}$ as
\begin{align}
&\hspace{-10pt}
\frac{P^{(2)}_{\nub_b\to\nu_a}}{L^2}
\equiv |\H{1}{a\bar b}|^2 
\nonumber \\
&
= \PCn{2}{a\bar b}
+\PAsn{2}{a\bar b} \sin \om_\oplus T_\oplus
+\PAcn{2}{a\bar b} \cos \om_\oplus T_\oplus
\nonumber \\
&\quad
+\PBsn{2}{a\bar b} \sin 2\om_\oplus T_\oplus
+\PBcn{2}{a\bar b} \cos 2\om_\oplus T_\oplus
\nonumber \\
&\quad
+\PDsn{2}{a\bar b} \sin 3\om_\oplus T_\oplus
+\PDcn{2}{a\bar b} \cos 3\om_\oplus T_\oplus
\qquad
\nonumber \\
&\quad
+\PFsn{2}{a\bar b} \sin 4\om_\oplus T_\oplus
+\PFcn{2}{a\bar b} \cos 4\om_\oplus T_\oplus .
\label{P2na sid}
\end{align}
Each of the nine amplitudes in this equation 
is a quadratic combination of the coefficients
$\gt^\ab_{a\bar b}$ and $\Ht^\al_{a\bar b}$
for Lorentz violation.
These combinations depend on the mass matrix
and also vary with the experimental scenario
through the neutrino energy 
and the direction of propagation.

The explicit forms of the amplitudes in Eq.\ \rf{P2na sid}
are somewhat lengthy and are omitted here.
However,
we can obtain compact expressions 
in terms of the amplitudes 
for the sidereal decomposition of $\H{1}{a\bar b}$,
which are defined in Eq.\ \rf{H1na sid}.
For the harmonics up to $2\om_\oplus$,
some calculation yields the results
\bea
\PCn{2}{a\bar b} &=&
|\Cn{1}{a\bar b}|^2
+\half |\Asn{1}{a\bar b}|^2
+\half |\Acn{1}{a\bar b}|^2
\nonumber \\ &&
\qquad
+\half |\Bsn{1}{a\bar b}|^2
+\half |\Bcn{1}{a\bar b}|^2 ,
\nonumber \\ 
\PAsn{2}{a\bar b} &=&
\Re \big[
2\Cn{1}{a\bar b}^*\Asn{1}{a\bar b}
+ \Acn{1}{a\bar b}^*\Bsn{1}{a\bar b}
\nonumber \\ &&
\qquad
- \Asn{1}{a\bar b}^*\Bcn{1}{a\bar b}
\big] ,
\nonumber \\ 
\PAcn{2}{a\bar b} &=&
\Re \big[
2\Cn{1}{a\bar b}^*\Acn{1}{a\bar b}
+ \Asn{1}{a\bar b}^*\Bsn{1}{a\bar b}
\nonumber \\ &&
\qquad
+ \Acn{1}{a\bar b}^*\Bcn{1}{a\bar b}
\big] ,
\nonumber \\ 
\PBsn{2}{a\bar b} &=&
\Re \big[
2\Cn{1}{a\bar b}^*\Bsn{1}{a\bar b}
+ \Asn{1}{a\bar b}^*\Acn{1}{a\bar b}
\big] ,
\nonumber \\ 
\PBcn{2}{a\bar b} &=&
2\Re\big[ \Cn{1}{a\bar b}^*\Bcn{1}{a\bar b} \big]
- |\Asn{1}{a\bar b}|^2
\nonumber \\ &&
\qquad
+ |\Acn{1}{a\bar b}|^2 ,
\eea
while for the harmonics at $3\om_\oplus$ and $4\om_\oplus$
we obtain
\bea
\PDsn{2}{a\bar b} &=&
\Re \big[
\Asn{1}{a\bar b}^*\Bcn{1}{a\bar b}
+\Acn{1}{a\bar b}^*\Bsn{1}{a\bar b}
\big] ,
\nonumber \\ 
\PDcn{2}{a\bar b} &=&
\Re \big[
\Acn{1}{a\bar b}^*\Bcn{1}{a\bar b}
-\Asn{1}{a\bar b}^*\Bsn{1}{a\bar b}
\big] ,
\nonumber \\ 
\PFsn{2}{a\bar b} &=&
\Re \big[
\Bsn{1}{a\bar b}^*\Bcn{1}{a\bar b}
\big] ,
\nonumber \\ 
\PFcn{2}{a\bar b} &=&
|\Bcn{1}{a\bar b}|^2
-|\Bsn{1}{a\bar b}|^2 .
\label{p2amps}
\eea
The structure of these equations
reflects the frequency dependence
in the sidereal decomposition \rf{H1na sid}
of $\H{1}{a\bar b}$.
For example,
the amplitudes 
$\PFsn{2}{a\bar b}$, $\PFcn{2}{a\bar b}$
for the fourth harmonic $4\om_\oplus$ 
of the probability $P^{(2)}_{\nub_b\to\nu_a}$
involve quadratic products of the amplitudes 
$\Bsn{1}{a\bar b}$, $\Bcn{1}{a\bar b}$
for the second harmonic $2\om_\oplus$ 
of $\H{1}{a\bar b}$,
as expected.

Comparable expressions for the CP-conjugate transition probability
$P^{(2)}_{\nu_b\to\nub_a}$
can readily be obtained
following the same procedure.
The results take the same form
as Eqs.\ \rf{P2na sid} and \rf{p2amps},
but with the index replacement
$\{a\bar b\}\to \{\bar ab\}$.

In the event that neutrino-antineutrino oscillations
are observed in nature,
the sidereal decomposition of the probability
$P^{(2)}_{\nub_b\to\nu_a}$ 
and its CP conjugate
offers a powerful approach to identifying
the relevant coefficients
$\gt^\ab_{a\bar b}$ and $\Ht^\al_{a\bar b}$.
Each experimental analysis separating 
the available sidereal harmonics 
would generate eight independent measurements,
with multiple experiments able to constrain
much of the available coefficient space.

\subsection{Oscillations conserving lepton number}
\label{sec lnc}

The analysis in the previous subsection demonstrates
that the detection of $\nu\mix\nub$ oscillations
is a unique signal for nonzero coefficients
$\gt^\ab_{a\bar b}$ and $\Ht^\al_{a\bar b}$.
However,
these coefficients also contribute at second order
to the more conventional $\nu\mix\nu$ 
and $\nub\mix\nub$ mixings.
For completeness,
we present the associated equations in this subsection.
Effects quadratic in the coefficients
$\aL^\al_{ab}$ and $\cL^\ab_{ab}$ also appear
at this order.
Inspection of Eq.\ \rf{P2an}
reveals that these contribute independently 
to the oscillation probabilities,
so we set them to zero here for simplicity.

The probabilities for $\nu\mix\nu$ and $\nub\mix\nub$ mixing 
are affected at second order by $\de h_{a\bar b}$
through its quadratic appearance in 
the quantities $\H{2}{ab}$ and $\H{2}{\bar a\bar b}$
defined in Eq.\ \rf{H2na}.
This produces sidereal variations 
at harmonics up to frequency $4\om_\oplus$.
We can therefore decompose $\H{2}{ab}$ as the sidereal expansion 
\begin{align}
&\hspace{-8pt}
\H{2}{ab} \equiv 
\sum_{c\bar d\bar ef}\M{2}{ab}{c\bar d\bar ef}
\de h_{c\bar d}\de h_{\bar ef}
\nonumber \\
&=\Cn{2}{ab}
+\Asn{2}{ab} \sin \om_\oplus T_\oplus
+\Acn{2}{ab} \cos \om_\oplus T_\oplus
\nonumber \\
&\quad
+\Bsn{2}{ab} \sin 2\om_\oplus T_\oplus
+\Bcn{2}{ab} \cos 2\om_\oplus T_\oplus
\nonumber \\
&\quad
+\Dsn{2}{ab} \sin 3\om_\oplus T_\oplus
+\Dcn{2}{ab} \cos 3\om_\oplus T_\oplus
\nonumber \\
&\quad
+\Fsn{2}{ab} \sin 4\om_\oplus T_\oplus
+\Fcn{2}{ab} \cos 4\om_\oplus T_\oplus .
\label{2ndcalh}
\end{align}
The nine amplitudes in the above expression
can be written as combinations of
the experiment-dependent factors $\M{2}{ab}{c\bar d\bar ef}$ 
and the five sidereal coefficients for 
the perturbative hamiltonian $\de h_{a\bar b}$
listed in Eq.\ \rf{h2Bc}.
For the amplitudes of the harmonics 
with frequencies $2\om_\oplus$ or less 
in the expansion \rf{2ndcalh},
we find the results
\bea
\Cn{2}{ab} &=& 
\sum_{c\bar d\bar ef}\M{2}{ab}{c\bar d\bar ef}\big[
\C{c\bar d}\C{\bar ef}\
\nonumber \\ &&
+\half \As{c\bar d}\As{\bar ef}
+\half \Ac{c\bar d}\Ac{\bar ef}
\nonumber \\ &&
+\half \Bs{c\bar d}\Bs{\bar ef}
+\half \Bc{c\bar d}\Bc{\bar ef} 
\big] , 
\nonumber \\
\Asn{2}{ab} &=& 
\sum_{c\bar d\bar ef}\M{2}{ab}{c\bar d\bar ef}
\big[
\C{c\bar d}\As{\bar ef}
+\As{c\bar d}\C{\bar ef}
\nonumber \\ &&
-\half \As{c\bar d}\Bc{\bar ef}
+\half \Ac{c\bar d}\Bs{\bar ef}
\nonumber \\ &&
+\half \Bs{c\bar d}\Ac{\bar ef}
-\half \Bc{c\bar d}\As{\bar ef}
\big] ,
\nonumber \\
\Acn{2}{ab} &=& 
\sum_{c\bar d\bar ef}\M{2}{ab}{c\bar d\bar ef}
\big[
\C{c\bar d}\Ac{\bar ef}
+\Ac{c\bar d}\C{\bar ef}
\nonumber \\ &&
+\half \As{c\bar d}\Bs{\bar ef}
+\half \Ac{c\bar d}\Bc{\bar ef}
\nonumber \\ &&
+\half \Bs{c\bar d}\As{\bar ef}
+\half \Bc{c\bar d}\Ac{\bar ef}
\big] ,
\nonumber \\
\Bsn{2}{ab} &=& 
\sum_{c\bar d\bar ef}\M{2}{ab}{c\bar d\bar ef}
\big[
\C{c\bar d}\Bs{\bar ef}
+\Bs{c\bar d}\C{\bar ef}
\nonumber \\ &&
+\half \As{c\bar d}\Ac{\bar ef}
+\half \Ac{c\bar d}\As{\bar ef}
\big] ,
\nonumber \\
\Bcn{2}{ab} &=& 
\sum_{c\bar d\bar ef}\M{2}{ab}{c\bar d\bar ef}
\big[
\C{c\bar d}\Bc{\bar ef}
+\Bc{c\bar d}\C{\bar ef}
\nonumber \\ &&
-\half \As{c\bar d}\As{\bar ef}
+\half \Ac{c\bar d}\Ac{\bar ef}
\big] .
\eea
For the remaining harmonics
in the expansion \rf{2ndcalh}
with frequencies $3\om_\oplus$ and $4\om_\oplus$,
the results for the amplitudes are 
\bea
\Dsn{2}{ab} &=& 
\sum_{c\bar d\bar ef}\M{2}{ab}{c\bar d\bar ef}
\nonumber \\ &&
\times 
\half\big[
\As{c\bar d}\Bc{\bar ef}
+\Ac{c\bar d}\Bs{\bar ef}
\nonumber \\ &&
+\Bs{c\bar d}\Ac{\bar ef}
+\Bc{c\bar d}\As{\bar ef}
\big] ,
\nonumber \\
\Dcn{2}{ab} &=& 
\sum_{c\bar d\bar ef}\M{2}{ab}{c\bar d\bar ef}
\nonumber \\ &&
\times 
\half\big[
-\As{c\bar d}\Bs{\bar ef}
+\Ac{c\bar d}\Bc{\bar ef}
\nonumber \\ &&
-\Bs{c\bar d}\As{\bar ef}
+\Bc{c\bar d}\Ac{\bar ef}
\big] ,
\nonumber \\
\Fsn{2}{ab} &=& 
\sum_{c\bar d\bar ef}\M{2}{ab}{c\bar d\bar ef}
\nonumber \\ &&
\times 
\half\big[
\Bs{c\bar d}\Bc{\bar ef}
+\Bc{c\bar d}\Bs{\bar ef}
\big] ,
\nonumber \\
\Fcn{2}{ab} &=& 
\sum_{c\bar d\bar ef}\M{2}{ab}{c\bar d\bar ef}
\nonumber \\ &&
\times 
\half\big[
-\Bs{c\bar d}\Bs{\bar ef}
+\Bc{c\bar d}\Bc{\bar ef}
\big] .
\qquad
\eea
Analogous expressions for the sidereal decomposition 
of the quantities $\H{2}{\bar a\bar b}$
and the resulting amplitudes 
can be obtained by substituting barred for unbarred indices 
and vice versa.

The second-order probability for neutrino-neutrino oscillations
inherit the same sidereal-frequency structure.
Introducing the expansion
\begin{align}
&\hspace{-10pt}
\frac{P^{(2)}_{\nu_b\to\nu_a}}{L^2}
\equiv -\Re \big((\S{0}{ab})^* \H{2}{ab}\big)
\nonumber \\
&
= \PCn{2}{ab}
+\PAsn{2}{ab} \sin \om_\oplus T_\oplus
+\PAcn{2}{ab} \cos \om_\oplus T_\oplus
\nonumber \\
&\quad
+\PBsn{2}{ab} \sin 2\om_\oplus T_\oplus
+\PBcn{2}{ab} \cos 2\om_\oplus T_\oplus
\nonumber \\
&\quad
+\PDsn{2}{ab} \sin 3\om_\oplus T_\oplus
+\PDcn{2}{ab} \cos 3\om_\oplus T_\oplus
\qquad
\nonumber \\
&\quad
+\PFsn{2}{ab} \sin 4\om_\oplus T_\oplus
+\PFcn{2}{ab} \cos 4\om_\oplus T_\oplus ,
\label{P2nn sid}
\end{align}
we find the nine corresponding amplitudes for the probability 
are given by the equations 
\bea
\PCn{2}{ab}&=& -\Re \big((\S{0}{ab})^* \Cn{2}{ab}\big) , 
\nonumber \\
\PAsn{2}{ab}&=& -\Re \big((\S{0}{ab})^* \Asn{2}{ab}\big) ,
\nonumber \\
\PAcn{2}{ab}&=& -\Re \big((\S{0}{ab})^* \Acn{2}{ab}\big) , 
\nonumber \\
\PBsn{2}{ab}&=& -\Re \big((\S{0}{ab})^* \Bsn{2}{ab}\big) , 
\nonumber \\
\PBcn{2}{ab}&=& -\Re \big((\S{0}{ab})^* \Bcn{2}{ab}\big) , 
\nonumber \\
\PDsn{2}{ab}&=& -\Re \big((\S{0}{ab})^* \Dsn{2}{ab}\big) , 
\nonumber \\
\PDcn{2}{ab}&=& -\Re \big((\S{0}{ab})^* \Dcn{2}{ab}\big) , 
\nonumber \\
\PFsn{2}{ab}&=& -\Re \big((\S{0}{ab})^* \Fsn{2}{ab}\big) , 
\nonumber \\
\PFcn{2}{ab}&=& -\Re \big((\S{0}{ab})^* \Fcn{2}{ab}\big) .
\eea
The probability for antineutrino-antineutrino oscillations 
can be found from the above equations by replacing
all indices $\{ab\}$ with $\{\bar a\bar b\}$.

The calculations in this subsection demonstrate
that searches for sidereal variations
in $\nu\mix\nu$ and $\nub\mix\nub$ oscillations
at the higher frequencies $3\om_\oplus$ and $4\om_\oplus$
can offer access to the coefficients 
$\gt^\ab_{a\bar b}$ and $\Ht^\al_{a\bar b}$
for Lorentz violation
without the need to study $\nu\mix\nub$ oscillations.
Moreover,
in addition to studies based 
on the above direct sidereal decompositions,
investigation of the CP and CPT asymmetries 
\rf{cpasym} and \rf{cptasym}
introduced in Sec.\ \ref{sec cpt}
provides another avenue for data analysis.
As before,
the time-averaged versions of these asymmetries
offer sensitivities to coefficients
that are challenging to detect 
in searches for sidereal variations.
In all these studies,
the second-order effects 
enter in conjunction with a factor of $L^2$,
so the large baselines associated with the experiments 
considered in Sec.\ \ref{sec ac}
imply that their intrinsic sensitivities to the coefficients
$\gt^\ab_{a\bar b}$ and $\Ht^\al_{a\bar b}$
are only mildly suppressed relative to 
the sensitivities to $\aL^\al_{ab}$ and $\cL^\ab_{ab}$.

\section{Summary}
\label{sec sum}

In this paper,
we study the effects of perturbative Lorentz and CPT violation 
on neutrino oscillations dominated by mass mixing.
The primary focus is on corrections 
arising from renormalizable operators 
for Lorentz violation within effective field theory.
In the neutrino sector,
these operators are controlled 
by SME coefficients for Lorentz violation denoted
$\aL^\al_{ab}$, $\cL^\ab_{ab}$,
$\gt^\ab_{a\bar b}$, and $\Ht^\al_{a\bar b}$.
They can affect conventional oscillations in
$\nu\mix\nu$ and $\nub\mix\nub$ mixing.
They also can induce $\nu\mix\nub$ mixing,
which violates lepton number. 

Using time-dependent perturbation theory,
a series expansion for the oscillation probabilities 
is derived in Sec.\ \ref{sec theory}.
To second order
in coefficients for Lorentz and CPT violation,
the probabilities for a nondegenerate mass spectrum
are presented in Eqs.\ \rf{zero1},
\rf{zero2}, \rf{P1na}, and \rf{P2an}.
At first order,
only the coefficients
$\aL^\al_{ab}$ and $\cL^\ab_{ab}$
contribute,
and lepton number is preserved.
Oscillations involving $\nu\mix\nub$ mixing 
appear at second order,
governed by the coefficients 
$\gt^\ab_{a\bar b}$ and $\Ht^\al_{a\bar b}$.

A key feature introduced by Lorentz and CPT violation 
is variations in the oscillation probabilities
with sidereal time.
The sidereal dependence arising from the coefficients 
$\aL^\al_{ab}$ and $\cL^\ab_{ab}$
is discussed in Sec.\ \ref{sec osc}.
It is described by the expansion \rf{P1sid},
which involves first and second harmonics
in the sidereal frequency.
At this order,
the amplitudes for each harmonic
are linear combinations of
$\aL^\al_{ab}$ and $\cL^\ab_{ab}$.
Data analyses using binning in sidereal time
can therefore measure these coefficients.

Section \ref{sec 3nu} 
addresses the methodology for data analyses
and provides illustrative estimates of numerical quantities
relevant for sidereal investigations 
in several long-baseline experiments.
The results are summarized
in Tables \ref{Table A} through \ref{Table D}.
For the three-generation case,
we demonstrate the procedure to identify
the relevant linear combinations of  
$\aL^\al_{ab}$ and $\cL^\ab_{ab}$,
using the K2K, MINOS, OPERA, ICARUS, NO$\nu$A, T2K, DUSEL, and T2KK
experiments as examples.
The two-flavor limit is also considered.
In this case,
the sidereal expansion is considerably simplified
and the oscillation probability 
takes the comparatively elegant form \rf{2nu prob}.

In addition to direct sidereal studies,
Lorentz and CPT violation can be sought 
through analysis of CP and CPT asymmetries 
in experimental data.
This topic is addressed in Sec.\ \ref{sec cpt}.
Suitable CP and CPT asymmetries
are defined in Eqs.\ \rf{cpasym} and \rf{cptasym}.
In the two-flavor limit,
these coincide and take 
the comparatively simple form \rf{2nu ratio} or \rf{mumuratio}.
Experiments running in both neutrino and antineutrino modes
can probe CP and CPT via this route.
Analyses along the lines proposed here 
could provide access to different combinations
of coefficients for Lorentz and CPT violation,
including ones that are challenging to detect
via studies of sidereal variations.

In Sec.\ \ref{sec gH},
we consider effects arising from nonzero coefficients 
$\gt^\ab_{a\bar b}$ and $\Ht^\al_{a\bar b}$.
Among the features is mixing between neutrinos and antineutrinos,
implying violations of lepton number.
These coefficients have no first-order perturbative effects.
Their dominant contributions arise at second order,
where the probabilities involve quadratic combinations
of $\gt^\ab_{a\bar b}$ and $\Ht^\al_{a\bar b}$.
This induces sidereal variations with harmonics
up to four times the sidereal frequency
in all three kinds of mixings,
$\nu\mix\nu$, $\nub\mix\nub$, and $\nu\mix\nub$.
The probabilities for neutrino-antineutrino mixing
are given in Eq.\ \rf{P2na sid},
while those for neutrino-neutrino mixing
take the similar form \rf{P2nn sid}.

Overall,
we find that the dominant effects
of renormalizable operators for Lorentz and CPT violation 
in the neutrino sector 
generate variations in oscillation probabilities
up to four times the sidereal frequency.
Subdominant perturbative effects may also 
offer useful information.
These higher-order perturbations 
cause sidereal effects at higher harmonics,
with signals suppressed compared to the ones discussed here.
We remark that other harmonics can also arise
from Lorentz-violating operators of nonrenormalizable dimensions
\cite{km1}.
A comprehensive SME-based study in analogy to that performed
for electrodynamics 
\cite{kmnonmin}
could establish the corresponding signals 
of Lorentz and CPT violation in neutrinos.

The results in this work
demonstrate that excellent sensitivity to Lorentz and CPT violation
is attainable by studying neutrino oscillations
with high energies and long baselines.
Our primary focus has been beam experiments,
where existing constraints 
\cite{lsndlv,sklv,minoslv}
span only a few percent
of the available coefficient space.
The procedures outlined in this work 
provide access to essentially all the coefficient space,
and moreover at sensitivities that can exceed 
the current ones by about two orders of magnitude.

An interesting direction for further work 
using a longer baseline
is a systematic investigation of perturbative effects 
of Lorentz and CPT violation on solar neutrinos,
for which day-night and annual signals
play a role analogous to sidereal effects
in beam experiments.
Future searches for Lorentz and CPT violation
using extreme baselines could also include 
studies of oscillations and dispersion for supernova neutrinos,
for which a sufficient population 
over a substantial solid angle 
would offer interesting sensitivity 
to a significant portion of the coefficient space.
A more speculative possibility using a cosmological baseline
would be the search for anisotropies 
in eventual observations of the cosmic neutrino background.
The maximal baseline makes this an ideal arena 
for studying low-dimension operators for Lorentz and CPT violation,
in analogy to the tight limits achieved 
on low-dimension operators in the photon sector 
using observations of the cosmic microwave background
\cite{cmb}.
In the meanwhile,
the long baselines involved in the many 
current and near-future beam experiments on the Earth 
imply impressive potential sensitivities to
the effects of Lorentz and CPT violation,
rivaling the best tests in other sectors of the SME.

\section*{Acknowledgments}

This work was supported in part
by the United States Department of Energy
under grant DE-FG02-91ER40661.

\end{document}